\documentclass[fleqn,usenatbib]{mnras}

\usepackage{newtxtext,newtxmath}
\usepackage{xcolor}

\usepackage[T1]{fontenc}
\usepackage{booktabs}
\usepackage{ulem}

\DeclareRobustCommand{\VAN}[3]{#2}
\let\VANthebibliography\thebibliography
\def\thebibliography{\DeclareRobustCommand{\VAN}[3]{##3}\VANthebibliography}

\usepackage{graphicx}	% Including figure files
\usepackage{amsmath}	% Advanced maths commands
\usepackage{color,ulem}  
\definecolor{deepblue}{rgb}{0,0,0.5}  
\definecolor{deepred}{rgb}{0.6,0,0}   
\definecolor{deepgreen}{rgb}{0,0.5,0} 
\definecolor{darkgreen}{rgb}{0,0.6,0} 

\def\ha{{\sc{H}}$\alpha$\/}
\def\habc{{\sc{H}}$\alpha_{\rm BC}$}

\def\hb{{\sc{H}}$\beta$\/}

\def\l{$\lambda$}

\def\eddr{$\lambda_{\rm Edd}$}
\def\l5100{L$_{\rm 5100}$}

\def\oiii{{[O\sc{iii}]}\/}

\def\ms{$M_{*}$}
\def\msun{M$_{\odot}$}
\title[Optical Properties of type I and II AGNs]{SDSS-IV MaNGA:The Incidence of Major Mergers in type I and II AGN Host Galaxies in the DR15 sample.}

\author[H. Hern\'andez-Toledo et al.]{Hern\'andez-Toledo H.M$^{1}$\thanks{E-mail: hector@astro.unam.mx}, Cortes-Su\'arez E.$^{1}$, V\'azquez-Mata J. A.$^{2}$, Nevin R.$^{3}$, \'Avila-Reese V.$^{1}$, 
\newauthor
Ibarra-Medel H.$^{4,5}$, Negrete C. A.$^{6}$ 
\\
$^{1}$Instituto de Astronomía, Universidad Nacional Autónoma de México, A.P. 70-264, 04510 CDMX, Mexico\\
$^{2}$Departamento de F\'isica, Facultad de Ciencias, Universidad Nacional Aut\'onoma de M\'exico, Ciudad Universitaria, CDMX, 04510, M\'exico\\
$^{3}$Fermi National Accelerator Laboratory, Batavia, IL 60510, USA  \\
$^{4}$Escuela Superior de F\'{\i}sica y Matem\'aticas, Instituto Polit\'ecnico Nacional, U.P. Adolfo L\'opez Mateos, C.P. 07738, Ciudad de M\'exico, M\'exico \\
$^{5}$Instituto de Astronom\'ia y Ciencias Planetarias, Universidad de Atacama, Copayapu 485, Copiap\'o, Chile\\
$^{6}$CONACyT Research Fellow -- Instituto de Astronomía, Universidad Nacional Autónoma de México, A.P. 70-264, 04510 CDMX, Mexico \\
}

\date{Accepted XXX. Received YYY; in original form ZZZ}

\pubyear{2015}

\begin{document}
\label{firstpage}
\pagerange{\pageref{firstpage}--\pageref{lastpage}}
\maketitle

\begin{abstract}

We present a study on the incidence of major mergers and their impact on the triggering of nuclear activity in 47 type I and 236 type II optically-selected AGN from the MaNGA DR15 sample. From an estimate of non-parametric image predictors ($Gini$, M$_{20}$, concentration (C), asymmetry (A), clumpiness (S), Sérsic index (n), and shape asymmetry($A_S$)) using the SDSS images, in combination with a Linear Discriminant Analysis Method, we identified major mergers and merger stages. We reinforced our results by looking for bright tidal features in our post-processed SDSS and DESI legacy images.
We find a statistically significant higher incidence of major mergers of 29\% $\pm$ 3\% in our type I+II AGN sample compared to 22\% $\pm$ 0.8\% for a non-AGN sample matched in redshift, stellar mass, color and morphological type, finding also a prevalence of post-coalescence (51\% $\pm$ 5\%) over pre-coalescence (23\% $\pm$ 6\%) merger stages.  
The levels of AGN activity among our massive major mergers are similar to those reported in other works using \oiii\ tracers. However, similar levels are produced by our AGN-galaxies hosting stellar bars, suggesting that major mergers are important promoters of nuclear activity but are not the main nor the only mechanism behind the AGN triggering. The tidal strength parameter $Q$ was considered at various scales looking for environmental differences that could affect our results on the merger incidence, finding non-significant differences. Finally, the H-H$\beta$ diagram could be used as an empirical predictor for the flux coming from an AGN source, useful to correct photometric quantities in large AGN samples emerging from surveys.

\end{abstract}


%%%%%%%%%%%%%%%%%%%%%%%%%%%%%%%%%%%%%%%%%%%%%%%%%%

%%%%%%%%%%%%%%%%% BODY OF PAPER %%%%%%%%%%%%%%%%%%

\section{Introduction}

Dynamical interactions and mergers, which promote gas infall, are among the main mechanisms suggested to drive the growth of supermassive black holes (SMBHs) in the nuclear regions of galaxies \citep[e.g.,][]{DiMatteo2005,Springel2005}. Tidal torques from major mergers can drive gas accretion for fueling both star formation \citep{Mihos1994,Mihos1996} and SMBH growth \citep{DiMatteo2005,Hopkins2005,Ellison2011,Koss2012,Treister2012,Satyapal2014}.   
Minor mergers are also thought among the most important mechanism for SMBH growth, simultaneously inducing morphological perturbations and stimulating star formation \citep[e.g.,][]{Noeske2007,Daddi2007,Cisternas2011,Kocevski2012,Kaviraj2014,Villforth2017}. 
The tight correlations between SMBH mass and galaxy properties suggest  co-evolution between them, with mergers possibly being the common cause of both SMBH accretion (with the consequent trigger of the AGN) and bulge mass buildup \citep{Hopkins2007}. Under these scenarios and depending on the initial conditions, galaxies hosting active galactic nuclei (AGN) could show to some degree those signatures of dynamical and morphological perturbations. 
However, other studies suggest that mergers do not necessarily play a dominant role in triggering AGN activity \citep{Grogin2015,Gabor2009,Cisternas2011,Kocevski2012,Schawinski2012,Treister2012,Simmons2013,Rosario2015}.

Alternatively, secular instabilities are another mechanism for driving gas accretion into SMBH in disk galaxies. These secular processes can be driven by bars or, in high redshift turbulent disks, by tidal friction of clumps, both also contributing to bulge growth \citep[e.g.,][and references therein]{Bournaud2016}. 

In the present work, we are particularly interested in identifying merger signatures in a carefully selected optical sample of type I and II AGN host galaxies from the Mapping Nearby Galaxies at Apache Point Observatory \citep[MaNGA;][]{Bundy2015, Yan2016a} DR15 survey \citep[][]{Aguado2019}. 
This approach is different from searching for the fraction of AGN hosts in samples of pair and/or post-merger galaxies. As discussed in \citet[][]{Ellison2019}, the latter focus on exploring whether or not mergers can trigger AGN, while our study will focus on the question of whether mergers are the dominant trigger of AGN or not.
However, our goal is fraught with difficulties associated to the recognition of mergers and the different merger stages; from early pre-coalescence to late post-mergers, to which different observational strategies and techniques are sensitive.
To capture a wide range of merger stages, those studies based on structural features and morphological distortions need of accurate schemes to associate observational signatures to those different stages.
Here, we adopt an accurate scheme recently proposed by \cite{Nevin2019,Nevin2023}. They used hydrodynamics simulations that cover a range of merger initial conditions coupled with dust radiative transfer codes to obtain highly realistic photometric properties. In that approach, they build mock observations of the simulated galaxies that allow them to create a classification of images, determining their accuracy and precision for identifying galaxy mergers of different mass ratios and interaction stages.  

In the recent years, Integral Field Spectroscopy (IFS) applied to large surveys has significantly improved the way of studying the galaxy properties and their connection to the hosted AGN \citep[e.g.,][]{Ibarra-Medel+16, Cano-Diaz+16,Sanchez2018,CanoDiaz2019, Gonzalez-Delgado+2017, Sanchez+2020,Aquino+2020,IbarraMedel2022}. In particular, for studies related to mergers and their connection to AGN, \citet{Jin2021} have exploited the IFS advantages by studying the role of AGN during galaxy interactions and how they inﬂuence the star formation by using a sample of 1156 galaxies in pairs or mergers from the MaNGA survey. Similarly, \citet{Steffen+2023} have compiled a sample of 391 spectroscopically confirmed galaxy pairs from the MaNGA survey to study the volume density of AGN and of dual AGN in galaxy pairs as a function of various projected separations.

We have taken advantage of (i) the spectral coverage of the MaNGA survey for a careful identification of our AGN samples \citep{Cortes2022} and of (ii) the spatial information contained in the SDSS r- band images of the host galaxies to estimate a series of optical morphological predictors (Gini, $M_{20}$, concentration, asymmetry, clumpiness, S\'ersic index, and shape asymmetry) that later are combined and interpreted via numerical simulations using Linear Discriminant Analysis (LDA) as described in \cite{Nevin2019}. We explore for: 
(i) the incidence (fraction) of major mergers in our samples of type I and II AGN galaxies; 
(ii) differences in this incidence between the combined type I and II AGN sample (hereafter type I+II) and a control non-AGN sample matched in stellar mass, morphology, color and redshift; and  
(iii) the level of AGN activity as a function of stellar mass for hosts with and without evidence of major mergers.

We also complement our analysis by carrying out a visual identification of tidal features in the type I and II AGN samples using our post-processing to the corresponding images from the Sloan Digital Sky Survey \citep[SDSS,][]{York2000,Stoughton2002} and Dark Energy Spectroscopy Instrument Legacy Imaging \citep[DESI images,][]{Dey2019}, up to $r$-band surface brightness limits of 25 and 26.7 mag arcsec$^{-2}$, respectively \citep{VazquezMata2022} .
Finally, we test various local and Large-Scale (LSS) environment indicators looking for possible differences in our samples that could affect our results on the incidence of major mergers.

This paper is organized as follows. We describe the MaNGA data, our AGN selection and their morphological and global physical properties in Section \S\S\ \ref{sec:data}. Section \S\S\ \ref{sec:3} describes the analysis and methods to infer the presence of mergers, merger stages and tidal features. The results obtained are presented in Section \S\S\ \ref{sec:results} with emphasis on the morphological content of the identified mergers, a statistical evaluation of the significance of the incidence of major mergers, the implications for the observed levels of nuclear activity, the alternative role of bars, and an evaluation of the impact that the local and LSS environment could have on our results. Section \S\S\ \ref{sec:discussion} puts into perspective our results, comparing with previous works and arguing about a dominant mechanism of AGN triggering in our samples. Finally, Section \S\S\ \ref{sec:conclusions} summarizes our main conclusions. We assume a standard Lambda-cold dark matter ($\Lambda$CDM) cosmology ($\Omega_M$ = 0.3, $\Omega_{\Lambda}$ = 0.7 and $H_0$ = 70 km s$^{-1}$ Mpc$^{-1}$). All magnitudes are in the AB system.

\section{The MaNGA survey: AGN galaxies and their optical global properties }
\label{sec:data}

The MaNGA survey \citep[SDSS-IV,][]{Blanton2017} has recently finished their Integral Field Spectroscopic observations of galaxies in the local universe ($z <$ 0.15) with a wavelength coverage from 3,600 to 10,400 \AA\ at a resolution ($\lambda/\delta \lambda$) of roughly 2000. The final DR17 \citep{Abdurro+22} includes the Data Reduction Pipeline \citep[DRP;][]{Law:2016aa,Law2021,Yan2016b} data products of 10,296 data cubes for MaNGA galaxies, with 10,145 of them having good data quality and no warning flags, yielding 10,010 unique targets (identiﬁed via their MANGAID) with a small number of repeated observations taken for cross-calibration purposes. The MaNGA data analysis pipeline (MaNGA DAP) is the package that analyzes the data produced by the MaNGA survey, however see also the implementation of the Pipe3D pipeline and of their data products \citep{Sanchez2016a,Sanchez2018,Sanchez2022}.

The final DR17 also reports the results of programs dedicated to AGN studies, such as the SPectroscopic IDentfication of ERosita Sources (SPIDERS) survey designed to provide an homogeneous optical spectroscopic follow-up of X-ray sources detected by eROSITA. A series of papers summarizing the results of the AGN programs are referred in \citet{Abdurro+22}.
Various other AGN catalogs were also compiled and analyzed throughout the entire MaNGA survey \citep[e.g.,][]{Rembold2017,Sanchez2018,Wylezalek2018,Wylezalek2020, Comerford2022}.
\citet{Comerford2020} presented the identifications of 406 AGNs out of 6261 galaxies observed at that stage in the MaNGA DR15 sample, collecting a series of multi-wavelength emission properties, dividing the AGNs into radio-quiet and radio-mode AGNs, and examining their galaxy star formation rates and stellar populations properties. More recently, \citet{Comerford2022} explored in more detail the available IFS data for a pilot sample of MaNGA galaxies looking for possible off-nuclear Seyfert regions, finding significant evidence of off-nuclear AGN signatures in their sample and showing that a more careful review of the whole DR17 MaNGA data could reveal a more complete census of AGNs missed by single fiber spectra.

\subsection{Type I and II AGN galaxies in MaNGA: the data}

In \citet[][hereafter Paper I]{Cortes2022} we have  carried out the accurate identification of 47 type I AGN in the MaNGA DR15 survey \citep[DR15;][]{Aguado2019}, containing at that stage 4636 galaxies. The selection method is based on the identification of the \ha\ broad component (\habc) by using a variant of the flux ratio method \citep[e.g.,][]{Oh2015}.  The higher signal-to-noise (S/N) ratio achieved in the integrated central (3 arcsec) spectra from the MaNGA data, avoided the need of a host galaxy subtraction.  
The method was further tested by using data from SDSS DR7, showing comparable results to other methods that identify type I AGN in the DR7 catalogs \citep[e.g.,][]{Stern2012,Oh2015,Liu2019}. The  \habc\ luminosity of their sample lies within the range  10$^{38}$ < L$_{\rm H_\alpha}$ < 10$^{44}$ erg s$^{-1}$,  with line widths log FWHM(\habc) $\sim$ 3-4, covering a range of Eddington ratio (\eddr, the ratio of the bolometric to Eddington luminosities) from $-5.15$ to 0.70 in logarithmic scale, with a few galaxies showing evidence of extended jet-like emission in radio wavelengths. We refer the reader to Paper I for a more detailed description and multiwavelength properties of the type I AGN sample. 

\begin{figure}
    \centering
    \includegraphics[width=\columnwidth]{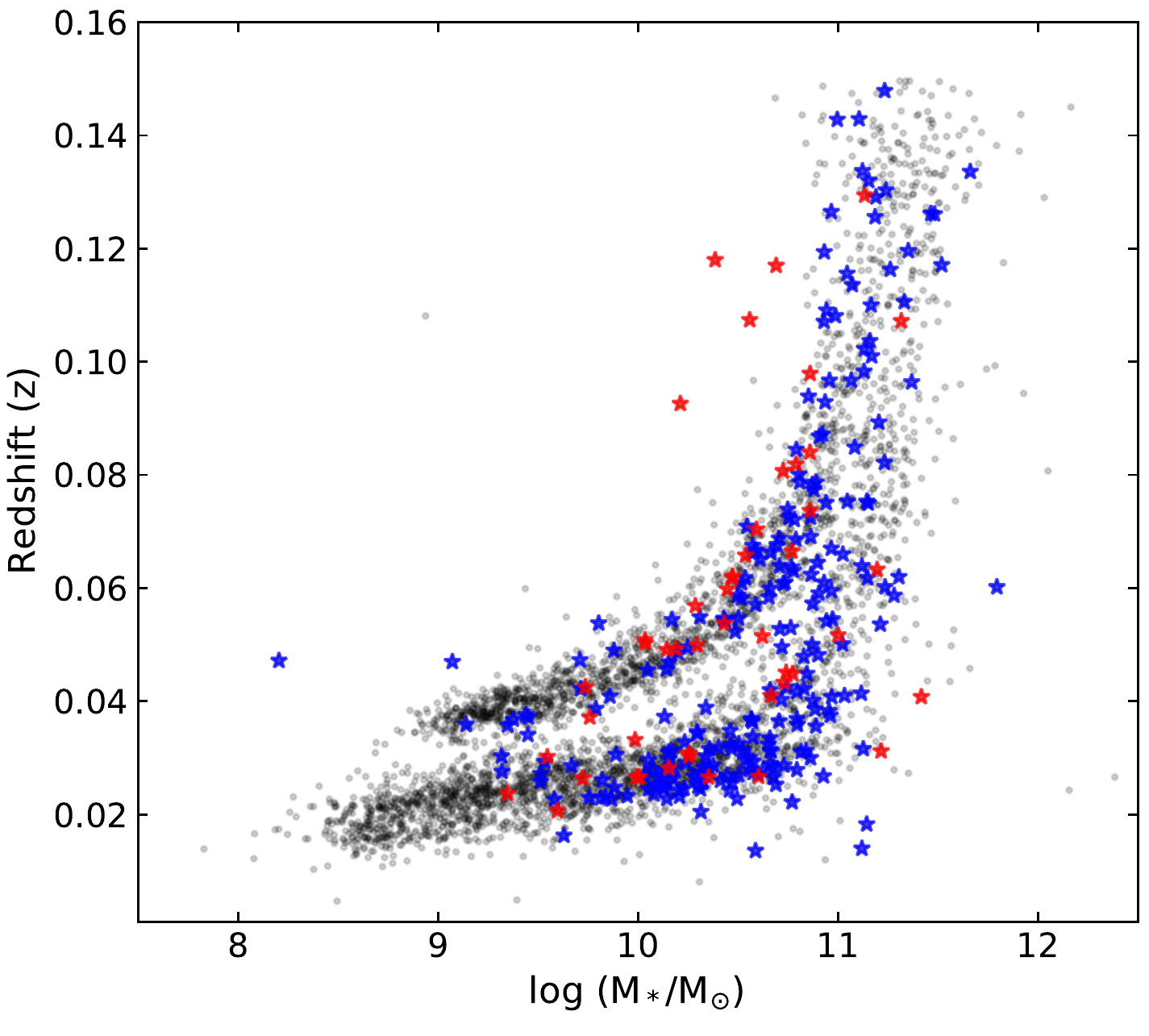}
    \vspace{-0.5cm}
    \caption{Distributions of type-I AGN (red stars), type-II AGN (blue stars) and non-AGN DR15 galaxies (gray dots) on the redshift vs stellar Mass diagram. Type I and II AGN populate almost equally ($\sim 50\%$) the two branches, corresponding to the MaNGA Primary$+$ and Secondary samples along the whole redshift interval.}
    
    \label{fig:Redshiftdistribution}
\end{figure}

\cite{Cortes2022} also looked for type II AGN in the MaNGA DR15 sample using the results from the Pipe3D data analysis software \citep{Sanchez2016a} adopting the criteria described in \cite{Sanchez2018}. They considered only galaxies located above the Kewley's lines \citep{Kewley2001,Kewley2006} in three independent BPT diagrams including the lines [O{\sc{iii}}], [N{\sc{ii}}] and [S{\sc{ii}}], and imposing the additional restriction on the \ha\ equivalent width, EW(\ha) > 1.5 \AA. While \cite{Sanchez2018} identified 98 type I and type II AGN candidates (in the MPL-5 containing $\sim$2700 galaxies), \cite{Cortes2022} obtained a final sample of 283 AGN: 47 type I and 236 type II (in the MPL-7 containing 4636 galaxies), which will be considered as the main sample for the present study.

Figure \ref{fig:Redshiftdistribution} shows the distribution of the 47 type I AGN (red symbols), the 236 type II AGN (blue symbols) and of a comparison sample of non-AGN MaNGA DR15 galaxies described below (gray symbols) on the redshift-stellar mass, $z-$\ms, diagram. Stellar masses have been estimated by using a color $(g-i)$ dependent mass-to-light ratio in the $i$-band following \cite{Taylor2011} and corrected from the contaminant AGN component as described in Appendix \ref{appendix:A}. Two separate sequences can be appreciated, corresponding to the MaNGA Primary$+$ and Secondary samples. Both type I and II AGN are well distributed between these two branches along the full redshift interval of the MaNGA survey.

\begin{table}
\begin{center}
\tabcolsep=0.15cm
\begin{tabular}{ccccccc}
\hline
\hline
\textbf{MaNGA-ID}	&	\textbf{g-i}	&	\textbf{z}	&	\textbf{M$_*$}	&	\textbf{Host} 	&	\textbf{Tidal}	&	\textbf{LD1}	\\
& & & & \textbf{Galaxy} & \textbf{Features} & \\
(1)	&	(2)	&	(3)	&	(4)	&	(5)	&	(6)	&	(7)		\\
\hline	\\
1-113712	&	1.005	&	0.081	&	10.725	&	SABab	&	No	&	8.736	\\
1-180204	&	0.927	&	0.052	&	10.621	&	SBbc	&	No	&	-6.715	\\
1-113405	&	1.070	&	0.042	&	9.773	&	Sa	    &   No	&	-4.379	\\
1-596598	&	1.109	&	0.027	&	10.352	&	SABa	&	No	&	-	\\
1-24092	    &	0.607	&	0.093	&	10.211	&	S0	    &	Yes	&	5.808	\\
1-24148	    &	1.113	&	0.028	&	10.129	&	Sb	    &	No	&	-4.565	\\
1-548024	&	1.154	&	0.129	&	11.133	&	SBb	    &	No	&	2.752	\\
1-43214	    &	0.831	&	0.118	&	10.792	&	S0a	    &   No	&	5.115	\\
1-121075	&	0.936	&	0.098	&	10.860	&	E$_{dc}$&	Yes	&	-0.271	\\
1-52660	    &	1.287	&	0.057	&	10.287	&	S0a	    &	Yes	&	6.562	\\
1-460812	&	1.195	&	0.067	&	10.770	&	Sa	    &	No	&	0.478	\\
1-523004	&	0.953	&	0.027	&	10.004	&	SBab	&	No	&	-2.308	\\
1-235576	&	0.964	&	0.070	&	10.593	&	SABa	&	No	&	0.509	\\
1-620993	&	0.856	&	0.030	&	10.254	&	SBbc	&	No	&	-4.647	\\
1-418023	&	0.937	&	0.024	&	9.348	&	S0	    &	No	&	-7.842	\\
1-256832	&	1.257	&	0.107	&	11.316	&	S0a	    &	Yes	&	2.114	\\
1-593159	&	1.100	&	0.045	&	10.773	&	SBa	    &	No	&	-0.161	\\
1-210017	&	1.116	&	0.043	&	10.730	&	SBb	    &	Yes	&	3.441	\\
1-90242	    &	0.360	&	0.030	&	9.546	&	E	    &	No	&	-7.916	\\
1-90231	    &	0.911	&	0.074	&	10.860	&	Sc	    &	No	&	-	\\
1-594493	&	1.174	&	0.031	&	11.215	&	E$_{dc}$&	Yes	&	-2.451	\\
1-95585	    &	1.125	&	0.063	&	11.198	&	SBc	    &	No	&	2.371	\\
1-550901	&	0.962	&	0.062	&	10.505	&	SBa	    &	No	&	4.276	\\
1-71974	    &	0.495	&	0.033	&	9.985	&	SBc	    &	No	&	6.417	\\
1-604860	&	1.108	&	0.052	&	11.002	&	E	    &	No	&	-1.504	\\
1-44303	    &	0.929	&	0.050	&	10.294	&	SBb	    &	Yes	&	-	\\
1-574519	&	1.030	&	0.049	&	10.143	&	Sa	    &	Yes	&	-3.849	\\
1-163966	&	0.775	&	0.027	&	10.603	&	SBa	    &	No	&	-0.809	\\
1-94604	    &	0.977	&	0.049	&	10.190	&	S0	    &	No	&	-3.938	\\
1-423024	&	1.077	&	0.062	&	10.488	&	SBab	&	No	&	-6.251	\\
1-174631	&	0.985	&	0.037	&	9.758	&	Sa	    &	No	&	-1.853	\\
1-149561	&	0.853	&	0.026	&	9.723	&	SBa	    &	No	&	-5.503	\\
1-614567	&	0.939	&	0.021	&	9.599	&	S0	    &	Yes	&	1.922	\\
1-210186	&	0.977	&	0.060	&	10.447	&	SBa	    &	No	&	-9.175	\\
1-295542	&	0.694	&	0.050	&	10.036	&	E$_{dc}$&	Yes	&	-2.430	\\
1-71872	    &	1.130	&	0.041	&	11.416	&	E$_{dc}$&	Yes	&	-3.672	\\
1-71987	    &	0.838	&	0.040	&	10.539	&	Sa	    &	Yes	&	1.784	\\
1-37863	    &	0.863	&	0.107	&	10.558	&	S0a	    &	Yes	&	3.128	\\
1-37385	    &	1.087	&	0.045	&	10.740	&	SBab	&	No	&	2.297	\\
1-37336	    &	1.123	&	0.084	&	10.859	&	Sab	    &	Yes	&	-3.856	\\
1-37633	    &	1.223	&	0.031	&	10.255	&	S0	    &	No	&	-1.610	\\
1-24660	    &	0.946	&	0.082	&	10.794	&	SBb	    &	No	&	-5.629	\\
1-574506	&	1.114	&	0.054	&	10.430	&	SB0a	&	No	&	-3.351	\\
1-574504	&	0.814	&	0.041	&	10.663	&	SBab	&	No	&	2.798	\\
1-298111	&	0.561	&	0.117	&	10.690	&	SBb	    &	Yes	&	-	\\
1-385623	&	0.632	&	0.051	&	10.035	&	S0a	    &	No	&	-5.875	\\
1-523211	&	1.088	&	0.027	&	9.993	&	S0	    &	No	&	-1.551	\\
\hline	
\hline
\end{tabular}
\end{center}
%\vspace{4pt}
\caption{Main properties of 47 galaxies with type I active nuclei according to our selection criteria \citep{Cortes2022}. Column (2) shows the $g$-band and $i$-band magnitudes from the NSA Sloan Atlas, corrected for the contaminant flux of a nuclear source after our Galfit 2D analysis (see Appendix \ref{appendix:A}). Column (3) the redshift obtained from NSA. Column (4) stellar masses from NSA catalog derived from Sersic fluxes. Column (5) the morphological Host Galaxy classification from \citet{VazquezMata2022} VAC. Column (6) tidal features detected in our image post-processing (see \S\S\ \ref{sec:decomposition}). Column (7) the Linear Discriminant Parameters to classify the galaxy as a major merger (see \S\S\ \ref{sec:LD1}).}
\label{tab:47type1}
\end{table}

Tables \ref{tab:47type1} and \ref{tab:type2general} present some global optical properties for the type I and type II AGN samples, respectively. Column (1) shows the MaNGA ID, column (2) the ($g-i$) color $K$- and extinction-corrected from NSA, column (3) the redshift coming from NSA, column (4) the stellar mass after adopting  corrections from the contaminant central source fluxes (see in Appendix \ref{appendix:A}),  column (5) the host galaxy morphological type from the MaNGA Visual Morphologies VAC DR17\footnote{\url{https://www.sdss.org/dr18/data_access/value-added-catalogs/?vac_id=80}}, column (6) a binary (Yes/No) identification of tidal features on the SDSS images (see \S\S\ \ref{sec:decomposition}), and column (7) the resulting LD1 parameter value after applying the Linear Discriminant Analysis (LDA) to our image predictors following \cite{Nevin2019} (see \S\S\ \ref{sec:LD1}).

\begin{table}
\begin{center}
\tabcolsep=0.15cm
\begin{tabular}{ccccccc}
\hline
\hline
\textbf{MaNGA-ID}	&	\textbf{g-i}	&	\textbf{z}	&	\textbf{M$_*$}	&	\textbf{Host} 	&	\textbf{Tidal}	&	\textbf{LD1}	\\
& & & & \textbf{Galaxy} & \textbf{Features} & \\
(1)	&	(2)	&	(3)	&	(4)	&	(5)	&	(6)	&	(7)		\\
\hline	\\
 12-84677	&	1.164	&	0.075	&	10.941	&	SAB0a	&	Yes	&	-2.919\\
 1-113651	&	1.169	&	0.071	&	10.546	&	SABab	&	No	&	-7.643\\
 1-547210	&	1.634	&	0.117	&	11.519	&	Sa	&	No	&	0.978\\
 1-547402	&	1.276	&	0.039	&	9.790	&	Sb	&	No	&	-24.579\\
 1-177493	&	1.276	&	0.108	&	10.989	&	SB0	&	Yes	&	-1.149\\
 1-177528	&	1.051	&	0.030	&	10.052	&	SAB0	&	No	&	-6.959\\
 1-180629	&	1.161	&	0.104	&	11.160	&	SBbc	&	No	&	-7.832\\
 1-180537	&	1.333	&	0.120	&	11.352	&	SBbc	&	No	&	-\\
 1-25554	&	1.032	&	0.027	&	10.282	&	SBa	&	No	&	-4.221\\
 1-595093	&	1.314	&	0.030	&	10.556	&	SABbc	&	No	&	-\\
 ...	&	...	&	...	&	...	&	...	&	...	&	...	\\
\hline	
\hline
\end{tabular}
\end{center}
\vspace{4pt}
\caption{Similar to table \ref{tab:47type1} Main properties of type II AGN selected according to our selection criteria. The complete data will be available in the online version.}
\label{tab:type2general}
\end{table}

Along the present paper various control samples were compiled and used for different purposes. A first control sample contains the full MaNGA DR15 sample, except for all identified type I and II AGN, useful to compare (i) the global physical properties of our AGN (\S\S\ \ref{sec:globalproperties}), and (ii) to test for possible effects of the local and Large-Scale environment in \S\S\ \ref{sec:results}. A second control sample was compiled by randomly selecting non-AGN galaxies per bin in stellar mass, color and redshift to match as much as possible the distributions of our type (I+II) AGN sample, to compare the frequency of bars. A third control sample, similarly matched in stellar mass and redshift, was compiled to compare the incidence of tidal features. Finally a control sample compiled by matching the morphological type, stellar mass, color, and redshift distributions of the AGN sample to test the significance of the frequency of major mergers identified with the LDA method in \S\S\ \ref{sec:Significance}. All the physical properties of our control samples were estimated and corrected in a similar way as those in the AGN samples.

\subsection{Morphological Type, Stellar Mass and Color}
\label{sec:globalproperties}

\begin{figure}
    \centering
    \vspace{-0.35cm}
    \includegraphics[width=\columnwidth]{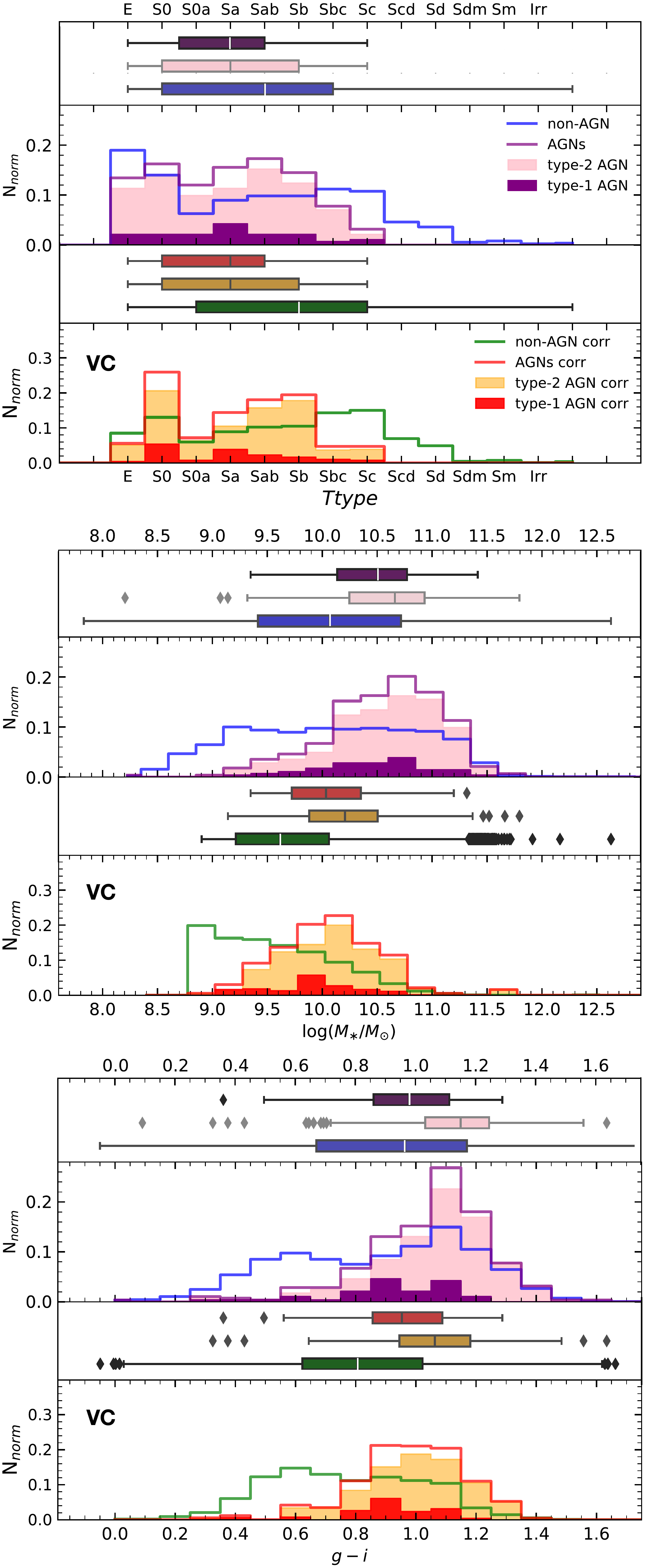}
    \caption{Morphological type (upper panels), stellar mass (middle panels) and ($g-i$) color distributions (lower panels) for non-AGN (blue and green solid lines), combined type I+II AGN (purple and red solid lines), the individual Type I (purple and red solid bars) and II AGN (pink and yellow solid bars). The upper sub-panel in each block corresponds to the observed quantities while the lower sub-panels stand for volume-corrected quantities (VC). On top of each panel a boxplot diagram shows the minimum, first quartile (Q1), median, third quartile (Q3), maximum and outliers of each distribution for non-AGN and type I and II AGN samples. }
    \label{fig:opical_properties}
\end{figure}

MaNGA galaxies were selected such that a roughly flat distribution in log $M_{i}$ was imposed, independent of morphology and color \citep{Wake2017}. This selection criterion does not guarantee a representative volume-complete sample either in \ms, color or morphology. Therefore, for some global quantities shown in the present work, we adopt volume completeness correction factors $f_m = f_m(M_\ast,g-r, z)$, that are applied to each galaxy in our AGN and control samples in order to recover representative quantities in the local volume. The details and methodology behind these corrections will be presented in Calette et al. (in prep). \citep[see also][]{Puebla2020,VazquezMata2022}. 

\begin{table*}
\begin{center}
%\tiny
\begin{tabular}{cccc|ccc|ccc|ccc}
\hline \hline

&\multicolumn{3}{c|}{Q1} & \multicolumn{3}{c|}{\textbf{Q2}} & \multicolumn{3}{c|}{Q3} & \multicolumn{3}{c|}{IQR} \\
\hline
&non-AGN &type I &type II &non-AGN &type I &type II &non-AGN &type I &type II &non-AGN &type I &type II \\
\hline

T Type & -1 & -0.5 & -1 & \textbf{2} & \textbf{1} & \textbf{1} & 4 & 2 & 3 & 5 & 2.5 & 4\\
T Type (VC) & 0 & -1 & -1 & \textbf{3} & \textbf{1} & \textbf{1} & 5 & 2 & 3 & 5 & 3 & 4\\
\hline
log(M$_*$) &  9.42 & 10.14 & 10.25 & \textbf{10.07} & \textbf{10.51} & \textbf{10.66} & 10.72 & 10.77 & 10.93 & 1.30 & 0.64 & 0.69 \\
log(M$_*$) (VC) & 9.21 & 9.72 & 9.88 & \textbf{9.62} & \textbf{10.04} & \textbf{10.21} & 10.06 & 10.35 & 10.50 & 0.85 & 0.63 & 0.62 \\
\hline
g-i & 0.67 & 0.86 & 1.03 & \textbf{0.96} & \textbf{0.98} & \textbf{1.15} & 1.17 & 1.11 & 1.24 & 0.50 & 0.25 & 0.21 \\
g-i (VC)  & 0.62 & 0.86 & 0.95 & \textbf{0.81} & \textbf{0.95} & \textbf{1.06} & 1.02 & 1.09 & 1.18 & 0.40 & 0.23 & 0.24\\

\hline 
\hline
\end{tabular}
\caption{Quartil values for the boxplots shown in Figure \ref{fig:opical_properties}. VC stands for volume-corrected distributions. The bold values are the Q2-median. }
\label{tab:quartile_hostboxplot}
\end{center}
\end{table*}

The morphological VC distributions of type I and II AGN are composed of S0-Sc galaxies, with a dominance of S0-Sb types. In both AGN samples we find only a small fraction of elliptical galaxies (11\% and 12\%, in type I and II, respectively) and no host galaxies with morphological types later than Sc. The overall distribution of our type (I+II) AGN sample shows a median $Ttype$ = 1 corresponding to Sa types, compared to a median $Ttype$ = 2 corresponding to Sab types for non-AGN hosts. This is consistent with results in other low-redshift AGN samples showing that they are preferentially hosted by early-type disk galaxies \citep{Ho1997,Kauffmann2003,Bruce2016,Sanchez2018,Kim2017,Kim2021}. 

Before volume-completeness corrections, the stellar mass distribution of the non-AGN sample is nearly flat, reflecting the $M_i$ selection criterion applied in the MaNGA survey and the peculiar selection of galaxies in the Color-Enhanced sample. Notice that the corresponding volume-corrected (VC) distributions are (i) trustworthy for galaxies with log(\ms/\msun) $>$ 8.8, and that (ii) these corrections change the mass and color distributions of the non-AGN sample and also (but to a lesser degree) those from the AGN samples to follow the local Stellar Mass Function (SMF) and slightly increasing the fraction of bluer galaxies, which typically are of lower masses (see Table \ref{tab:quartile_hostboxplot}). 

Type (I+II) AGNs are hosted in galaxies with median VC stellar masses log(\ms/\msun) = 10.04 and 10.21 respectively, compared to log(\ms/\msun) = 9.62 for non-AGN. The ($g-i$) color, an approximate tracer of the recent ($\lesssim$ 1 Gyr) mean SFR history shows median VC $(g-i)$ = 0.95 and 1.06 for type I and type II AGNs, respectively, while that of non-AGN hosts is ($g-i$) = 0.81. These differences, more noticeable in the volume-corrected distributions, show that 
galaxies hosting type (I+II) AGN tend to be more massive and redder than those in the non-AGN control sample, although with large overlaps in the distributions. A Kolgomorov-Smirnov test shows that the mass and color distributions of type (I+II) AGN and non-AGN galaxies are different at the significance level $p = 5\times 10{-16}$, while a T-test shows that the mean color and stellar mass in the AGN sample are greater than those in the control sample with a similarly high significance level. In Sect. \ref{sec:discussion} we discuss some implications of these differences.

\section{Identification of Major Mergers, Merger Stages and Tidal Features}
\label{sec:3}

\subsection{Major mergers identification}
\label{sec:LD1}

Several works in the literature have proposed methods to identify signatures of mergers in galaxies. Among them we mention those using non-parametric image predictors such as the $Gini-M_{20}$ or the CAS (Concentration-Asymmetry-Clumpiness) \citep{Lotz2004,Conselice2003}. Although very useful, their prediction ability is limited by different merger initial conditions, such as mass ratio, gas fraction, merger stage and the merger observability timescale \citep[e.g.,][]{Lotz2008,Lotz2011}. Recently, \cite{Nevin2019} proposed a merger identification scheme that simultaneously uses several image predictors like $Gini$, $M_{20}$, Concentration ($C$), Asymmetry ($A$), Clumpiness ($S$), S\'ersic index ($n$), and Shape Asymmetry ($A_S$) exploiting their individual prediction ability and combining them to generate a more robust classification using Linear Discriminant Analysis (LDA). They trained separate minor merger (mass ratio of the galaxies 1:5 and 1:10) and major merger (mass ratio of the galaxies 1:2 and 1:3) classifiers using a suite of mock images from detailed hydrodynamic simulations of merging galaxies, including AGN. By combining the coefficient values of the different predictors, they are able to accurately identify merging galaxies over a range of mass ratios, gas fractions, viewing angles, and merger stages. In the present work, the LDA method of \cite{Nevin2019} will be used to identify only major mergers.

The simulated images in \cite{Nevin2019} were SDSS-ized introducing the SDSS noise and background characteristics and convolving to the seeing limit of the SDSS survey to gain more direct applicability of their merger classification scheme to the complete SDSS image survey. Thus, for consistency the image predictors for our AGN and non-AGN control samples were estimated from the $r$-band SDSS images using a software from \cite{Nevin2023} that includes the Python package \textit{statmorph} \citep{Rodriguez-Gomez2019}, with various modifications, to prepare the images (background-subtraction, masking and segmentation) prior to compute the source properties following the procedures described in \cite{Abraham2003,Conselice2000,Conselice2003,Lotz2004,Lotz2008,Pawlik2016,Sersic1963}.

\cite{Nevin2019} optimized the applicability of the LDA method in terms of the SDSS image properties, finding it more appropriate for images with a minimum  S/N ratio $\sim$ 2.5 per pixel on the $r$-band, corresponding to features with surface brightness $\sim$ 25 mag arcsec$^{-2}$, objects brighter than 17 mag and within the redshift $z < 0.5$ range. Except for very few cases, all the $r$-band SDSS images corresponding to our AGN and control samples, satisfy those requirements for its robust application (see also Appendix \ref{appendix:B}).

\begin{figure*}
    \centering
    \includegraphics[scale=0.7]{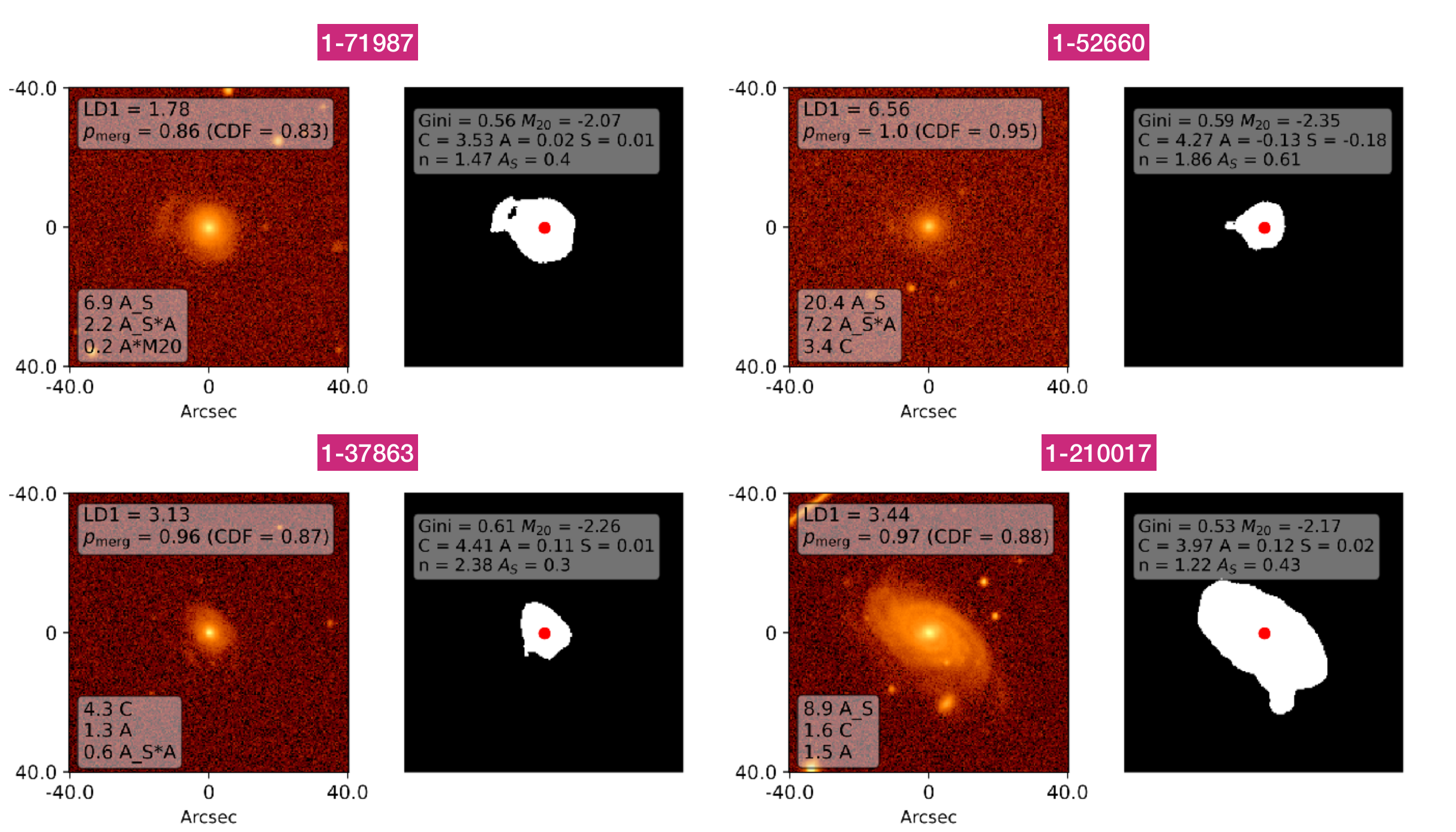}
   
    \caption{Diagnostic images (left panels) and segmentation maps (right panels) for a selection of galaxies classified with high probabilities of being major mergers. The diagnostic diagrams include the LD1 value and corresponding $p_{\mathrm{merg}}$ and CDF value for each individual galaxy in the inset box in the left top as well as the leading coefficients and corresponding predictors in the bottom left inset box. Finally, the predictor values are provided in the upper right inset panel. We also include the MaNGA ID (in purple) for each galaxy. Most of these galaxies exhibit disturbed features such as shells (upper left) or companion galaxies (lower right) which can be seen in the segmentation maps. These features contribute to high shape asymmetry values, which lead to large LD1 values.}
    \label{fig:segmentation_maps}
\end{figure*}

Figure \ref{fig:segmentation_maps} illustrates some procedures behind the estimate of the image predictors and the application of the LDA method. The left panel shows an $r$-band image in false-color from which the corresponding segmentation map is built (right hand panel). The galaxy morphology tool \texttt{statmorph} (\citealt{Rodriguez-Gomez2019}) and the \texttt{photutils} package were used to define the segmentation map using a 1.5$\sigma$ threshold above the background. The insets emphasize the most influential terms in the LDA classification; namely, the  Linear Discriminant Parameter (LD1), the major merger probability ($p_{\mathrm{merg}}$) and the Cumulative Distribution Function (CDF) value, which compares the $p_{\mathrm{merg}}$ value of each individual galaxy to that of all galaxies in SDSS. The inset on the segmentation map show the values of the corresponding non-parametric morphological predictors that were combined and that drive the major merger classification. Tables \ref{tab:type1imaginpredictors} and \ref{tab:type2imaginpredictors} in Appendix \ref{appendix:B} report the results of the image predictors estimates from the SDSS r -band images for our type I and II AGN samples.\footnote{The corresponding Table for the full non-AGN control sample will be presented by (Vazquez-Mata et al. 2023 in prep) as an update to the MaNGA morphological VAC}.

Once the image predictors were estimated for our AGN and control samples, the next step is to apply the LDA method to optimize the  major merger classification. According to \cite{Nevin2019}, the decision boundary value of the LD1 parameter for the major merger combined runs is 1.16, thus all galaxies with $LD1 > 1.16$ will be classified as major mergers. Alternatively, in terms of merging probability ($p_{\mathrm{merg}}$), all galaxies having $p_{\mathrm{merg}} > 0.76$ will be considered as major mergers (See Tables \ref{tab:47type1} and \ref{tab:type2general}, Column (7)). As previously mentioned, in the present work, we will consider the classifiers from the combined major merger simulations, and thus the classification reported here is related only to major mergers.

\subsubsection{Major Merger Stages} 
\label{sec:merging-stages}

There is compelling evidence showing that merger-induced AGN activity is more important at more advanced stages of the merger process \citep{Ellison2013,VanWassenhove2012,Bickley2023}. High-resolution hydrodynamical simulations predict that AGN activity generally increases with more advanced merger-stages but particularly as a pair coalesces into a post-merger stage \citep{VanWassenhove2012,Capelo2017}. More recently, \cite{Byrne-Mamahit2023} investigated the accretion rates of SMBHs in post-merger galaxies drawn from the IllustrisTNG simulation finding accretion rates $\sim$ 1.7 times higher than in control samples, also finding that the presence of simultaneous enhancements in either the star formation and SMBH accretion rates depends on both the mass ratio of the merger and on the gas mass of the post-merger galaxy.

Although the LDA method was originally designed for the entire duration of the merger process (from early to post-coalescence stages), more recently \cite{Nevin2023}, following previous theoretical and observational work, has further refined the LDA classification into different stages to gain insight on the time-dependent evolutionary processes in mergers. That classification method is able to assign each galaxy a probability of being in a merging stage. They divided their classification into pre and post-coalescence stages to match the methodology of cosmological merger identification schemes \citep[e.g.,][]{Hani2020,Bickley2021} and further divided the pre-coalescence classification into early and late stages to roughly match the stages in \cite{Moreno2015} and \cite{Pan2019} of first pericentric passage and apocenter (early) and final approach (late). They also implemented a sliding timescale for the definition of the post-coalescence stage, using a time cutoff of 0.5 Gyr after colaescence and then additionally implementing a time cutoff of 1 Gyr, following \cite{Bickley2021} who found that the morphology of Illustris TNG galaxies is disturbed for up to 2.5 Gyr following a merger.

\subsection{Tidal features}
\label{sec:decomposition}

Observations as well as numerical simulations have shown how baryonic matter can trace merging events through the formation of discernible tidal features \citep{vanDokkum2005,Tal2009,Kaviraj2010,Sheen2012,Kim2013,Hong2015,Mancillas+2019}. We present the results of a visual identification of bright tidal features in our AGN and control samples by using our $r$-band image post-processing from the SDSS and DESI surveys. Although our search is not detailed and far from complete \citep[for related results, see][]{VazquezMata2022}, this tidal census can be very useful to understand the morphological nature of the major mergers identified with the Linear Discriminant Analysis (LDA) method and their relation to other galaxy properties.

In \cite{VazquezMata2022} both the $r$-band SDSS and DESI images of our AGN and control samples were post-processed by preserving their native pixel scales of 0.396 ''/pix and 0.262''/pix, with a typical image quality FWHM (arcsec) = 1.4 and 1.3 respectively. Since both surveys also provide depth maps for each stacked image in each band, we have estimated the $r$-band 5$\sigma$ depths at the outskirts of each AGN host, finding average surface brightness limits $\sim$ 25 and 26.7 mag arcsec$^{-2}$ respectively. The residual images of the post-processing catalogue from the DESI Legacy Survey were also retrieved and included in our identification mosaics, providing valuable information of morphological features in the inner and outer regions of galaxies. For more details on the residual images, and on The Tractor package see \citet[][]{Dey2019}.   

\begin{figure}
    \centering
    \includegraphics[width=\columnwidth]{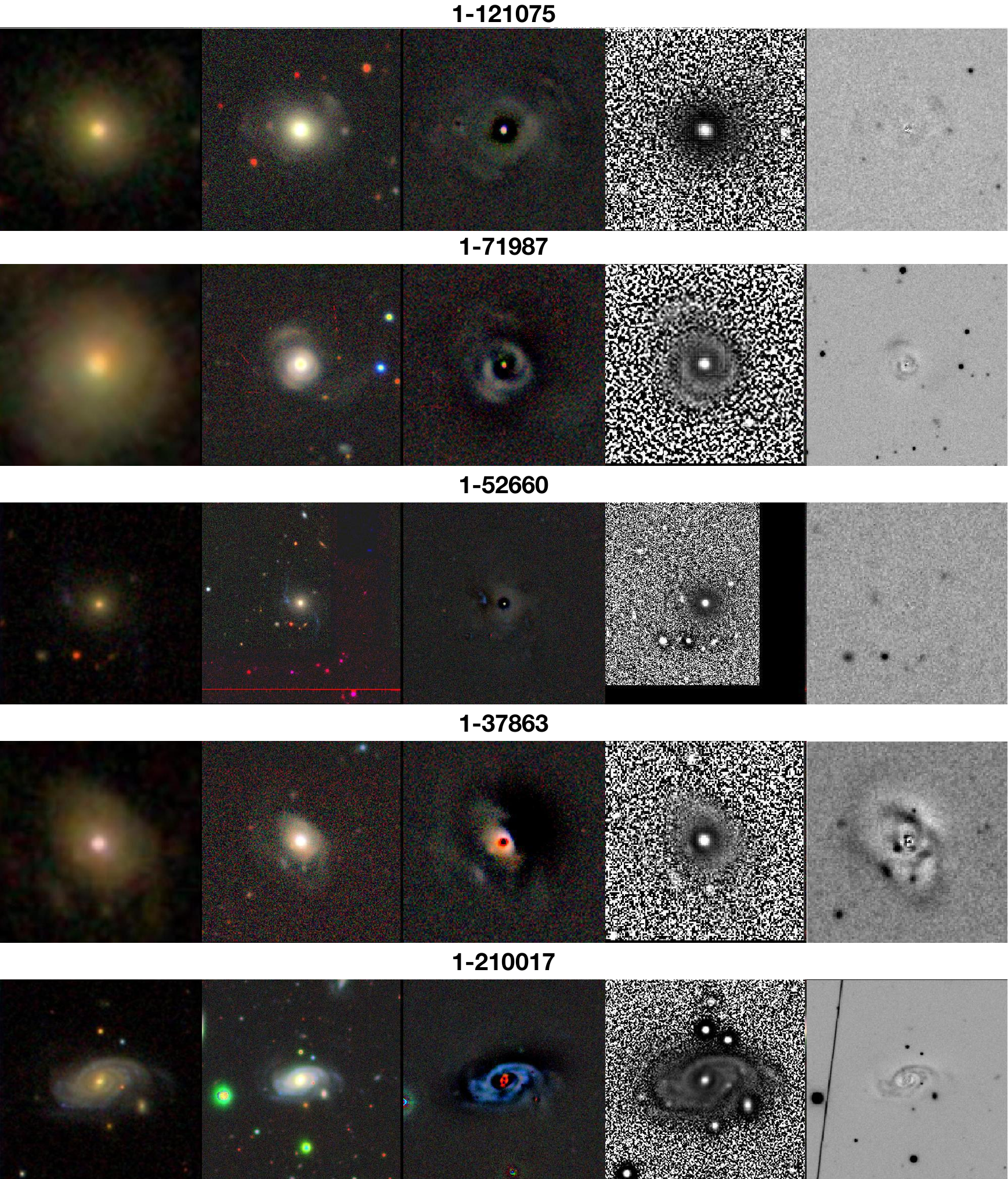}
   
    \caption{SDSS, DESI Legacy Post-processed Images and Galfit Residuals for the Major Merger Candidates. From left to Right; SDSS $gri$ image, DESI $grz$ image, DESI residual after subtracting a set of parametric light profiles, the DESI filter-enhanced $r$-band image and the residual $r$-band image after subtracting a 2D bulge/disk/bar/nuclear source. The presence of shells are evidenced in these images. }
    \label{fig:mos-typeI}
\end{figure}

\begin{figure}
    \centering
    \includegraphics[width=\columnwidth]{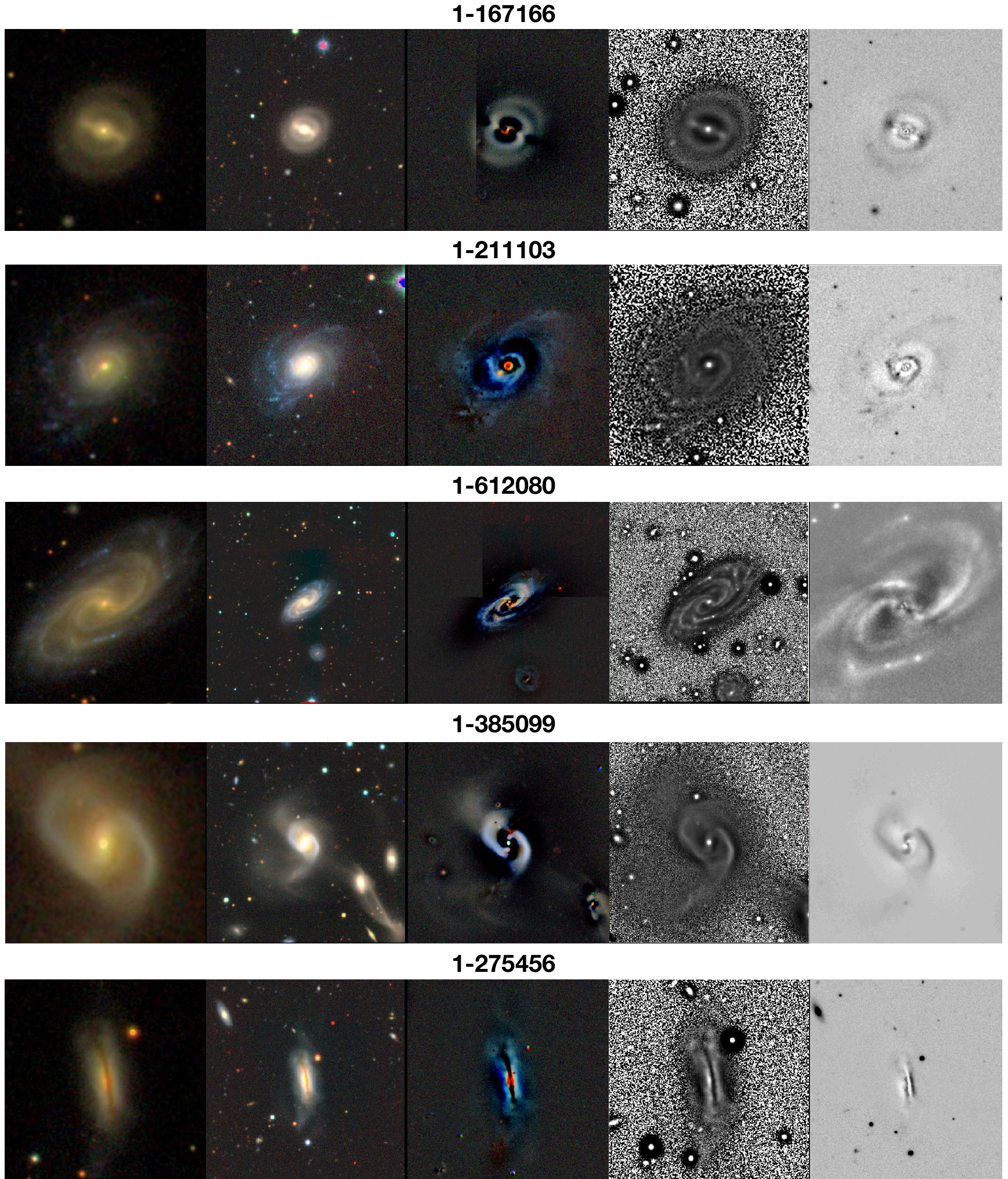}
   
    \caption{Same as figure \ref{fig:mos-typeI} but for type II AGN hosts.
    }
    \label{fig:mos-typeII}
\end{figure}

Figure \ref{fig:mos-typeI} summarizes the results of our image post-processing in the form of mosaics for a few examples of type I AGN hosts. From left to right; the $gri$ SDSS composite image, the $grz$ DESI composite image, the PSF-convolved residual DESI image (after subtracting the best 2D surface brightness model), the DESI filtered-enhanced $r-$band image and the residual $r$-band image (after subtracting our best Galfit 2D surface brightness model). Similar mosaics for type II AGN hosts are shown in Figure \ref{fig:mos-typeII}. Notice the variety of morphological types associated to our AGN samples.

\section{Results}
\label{sec:results}

\subsection{Major mergers and merger stages with the LDA method}
\label{sec:major-mergers}

The numerical values of the image predictor parameters (Sect. \ref{sec:LD1}) for our AGN samples, are reported in Tables \ref{tab:type1imaginpredictors} and \ref{tab:type2imaginpredictors}. The resulting LD1 major merger classifier parameters are presented in Column (7) of Tables \ref{tab:47type1} and \ref{tab:type2general}. For type I AGN, the LDA method identified 15 major mergers (LD1 > 1.16) out of 43  AGN hosts (35\% $\pm$ 7\%) with good quality data ($m_r < 17$ and $S/N > 2.5$), from which 13 belong to S0-Sb morphological types and two to Sc type. Neither of the six elliptical galaxies in this sample were identified as a major merger. For type II AGN, 55 major mergers were identified out of 214 hosts (26\% $\pm$ 3\%) with good quality data, from which 43 belong to S0-Sb types and 5 to Sbc type. Error estimates correspond to the standard deviation as computed from binomial statistics \footnote{$\sigma$ = $\sqrt{npq}$, where $n$ is the total sample, $p$ the probability of success and $q = 1 - p$.}
This time, 7 elliptical galaxies were identified as major mergers. Our results show a dominant fraction of early type disks ($<$ Sb types) among the identified mergers, a result expected from the abundance of early type disk galaxies among our AGN samples (see Section \ref{sec:globalproperties}). Notice however, that other observational works using different identification schemes like the visual identification of pairs of merging galaxies from the Galaxy Zoo project  \citep[e.g.,][]{Darg2010}, also show a prevalence of early-type spirals over elliptical galaxies, similar to our results.
Although the LDA method was trained mostly upon disk-dominated galaxies, a recent analysis of its performance by \cite{Nevin2023} shows its robustness in identifying mergers across a wide range of morphologies, stellar masses, and redshifts. Therefore, a bias related to the morphological content of the identified mergers is not expected.  

We next proceed to identify the merger stages adopting the hierarchical method presented in \cite{Nevin2023}. To that purpose we compared the $p_{\rm merg}$ (50th percentile) probability of being in a merging stage (pre-coalescence or post-coalescence), choosing the higher $p_{\rm merg}$ value. Since the 16th and 84th percent values associated to each $p_{\rm merg}$ were also reported, an interquartil (84th - 16th) difference is used as a gross confidence interval to distinguish between pre-coalescence and post-coalescence merger stages.

For type I AGN, the hierarchical method finds 66\% $\pm$ 5\% in the post-coalescence stage and 20\% $\pm$ 6\% in the pre-coalescence stage with 13\% of unclassified cases.  For type II AGN, the method finds 46\% $\pm$ 4\% in post-coalescence stage and 24\% $\pm$ 5\% in the pre-coalescence stage, this time with 29\% of unclassified cases. The results for the combined type (I+II) AGN sample are 51\% $\pm$ 5\% in the post-coalescence stage and 23\% $\pm$ 6\% in the pre-coalescence stage, with 26\% of unclassified cases. The unclassified cases refer to those having an outlier predictor flag in their photometric properties, meaning that the predictor values are outside of the predictor values of the simulated training set. Some of them have also low S/N flags. Thus they are not included in the classification.

If the interquartile difference is adopted as a confidence interval around each $p_{\rm merg}$ value, then we can cleanly separate merger stages only for about half of the mergers; 13\% in pre-coalescence and 26\% in post-coalescence for the combined type (I+II) AGN sample, with the other half being not-distinguishable within such broad interquartil difference.   
In any case, the hierarchical method indicates approximately \textit{a factor of 2 prevalence of post-coalescence merger stages over pre-coalescence stages among the major mergers identified with the LDA method in our AGN samples. }

For each predicted merger, we further used our image mosaics in combination with the DESI Legacy Survey Sky Browser to carry out and independent visual classification of the merger stages. We looked for galaxies within 1:2 and 1:3 the size of each target inspecting a wide field of view around each target, retrieving information on their radial velocities to confirm a physical association within 600 km s$^{-1}$. We named a first category as separated or pre-merger stage, referring to systems having separations grater than two apparent diameters of the major merger target. A second category was named as advanced/post-merger stage, referring to galaxies well within one diameter of the major merger target, galaxy cores not settled yet sharing a common diffuse light envelope, and galaxies already settled with the appearance of a single galaxy showing evidence of tidal features.  

Our visual identification yields 60\% $\pm$ 5\% in post-coalescence stage and 40\% $\pm$ 5\% in pre-coalescence stage for type I AGN mergers. For type II AGN mergers, 68\% $\pm$ 5\% are in post-coalescence stage and 32\% $\pm$ 5\% in pre-coalescence stage. The results for the combined type (I+II) AGN sample are 65\% $\pm$ 5\% in post-coalescence stage and 35\% $\pm$ 5\% in pre-coalescence stage.
As with the hierarchical method, this analysis shows a higher incidence of post-coalescence over pre-coalescence merger stages among our major mergers.

A comparison of these results indicates that (i) there is no overlap of the  percentages (within the quoted uncertainties) predicted by the LDA and visual methods. However, they are in qualitative agreement, both finding a higher number of post-mergers over pre-mergers. (ii) For type I AGN mergers, these percentages are not far from one another. For type II AGN mergers, the higher incidence of unclassified cases makes the comparison more difficult. However, notice that (iii) the fraction of (pre + post coalescence) coincidences between the LDA and visual methods is relatively high, reaching 66\% in type I AGN and 57\% in type II AGN, with 59\% for the combined type (I+II) AGN sample.

\subsection{Tidal features}
\label{sec:visual}

For type I AGN, we find visual evidence of bright tidal features in the SDSS images in 14 out of 47 host galaxies (30\% $\pm$ 7\%). For type II AGN, such evidence is found in 73 out of 236 host galaxies (31\% $\pm$ 3\%). In addition, when using the deeper images from the DESI survey, we find tidal features in 16 out of 47 host galaxies (34\% $\pm$ 7\%) in type I AGN, while for type II AGN, these are found in 83 out of 236 host galaxies (35\% $\pm$ 3\%). The occurrence of tidal features and the morphological content of their corresponding host galaxies are summarized in Table \ref{tab:tidalfeat}. Given that the LDA method was optimized upon the SDSS image properties, for consistency we report in Column (6) of Tables \ref{tab:47type1} and \ref{tab:type2general} a binary (Yes/No) ﬂag indicating a (positive/negative) detection of tidal features in type I and type II AGN hosts after inspecting the SDSS images.

\begin{table}
\centering
\begin{tabular}{c|cccc}
\hline 
\hline
\multicolumn{5}{c|}{Type I and II AGN Hosts with Tidal Features} \\
\hline
 & \multicolumn{4}{c|}{\textbf{type I AGN}} \\
\hline
Images & Early & Early Disk & Late Disk & Total  \\
r -band & (E) & (S0-Sb) & (Sbc-Sc) & (out of 47) \\
\hline
SDSS  & 3 & 11 & - & 14 (30$\pm$7)\% \\
DESI  & 4 & 12 & - & 16 (34$\pm$7)\%\\
\hline
\hline
 & \multicolumn{4}{c|}{\textbf{type II AGN}} \\
\hline
Images & Early & Early Disk & Late Disk & Total\\
 & (E) & (S0-Sb) & (Sbc-Sc) & (out of 236)  \\
\hline
SDSS  & 11 & 60 & 2 & 73 (31$\pm$3)\% \\
DESI   & 16 & 65 & 2 & 83 (35$\pm$3)\% \\

\hline 
\hline
\end{tabular}
\caption{The fraction of tidal features in type I and type II AGN samples and their morphological content.}
\label{tab:tidalfeat}
\end{table}

Our results using the SDSS images indicate an almost factor of two higher incidence of tidal features in the combined type (I+II) AGN sample when compared to 16\% $\pm$ 0.7\% found for galaxies in a non-AGN control sample matched in stellar mass and redshift (see also \citealp{VazquezMata2022}), that increases to a factor slighly higher than two when using the DESI images.

It is worth mentioning that while most of the apparent tidal features in the SDSS images were confirmed with the DESI images, some were not, finding instead, that they are part of other structural components (outer arms or rings). Other subtle features were also identified in the deeper DESI images yielding thus a slightly higher fraction of tidal features than in the SDSS images.
 
\citet{Nevin2019} assessed the accuracy of the LDA method finding that the SDSS-based results show a higher false negative rate than false positive rate, meaning it is more likely to miss mergers than missclassify non-merging galaxies as mergers. Since we lack a more detailed analysis of the performance of the LDA method trained on DESI images, we cannot say for sure, but we suspect that the LDA method trained on the deeper DESI images would have a lower false negative rate. In particular, it might be more sensitive to mergers with faint tidal tails and/or higher redshift mergers (though this is beyond the scope of this work).

\subsection{LDA major mergers versus tidal features}

As described in \S\S\ \ref{sec:LD1}, the LDA method was not trained on the basis of a visual identification of tidal features, so we expect that major mergers selected with the LDA method do not necessarily capture the evidence on the wide variety of tidal features found. A cross-match of Tables \ref{tab:47type1} and \ref{tab:type2general} confirms that not all galaxies qualifying as major mergers with the LDA method (LD1 > 1.16) show tidal features, nor all galaxies with tidal features are identified as major mergers with the LDA method. For type I AGN, on one side, the LDA method identifies 15 major mergers out of 43 hosts (35\% $\pm$ 7\%) with good quality data. On the other side, 14 out of 47 hosts (30\% $\pm$ 7\%) show evidence of tidal features. In this case, more than half of the major mergers show tidal features (9 coincidences), all nine belonging to S0-Sb types. Among the 6 ellipticals in this sample, 3 show evidence of tidal features, but none of them was identified as a major merger with the LDA method. Similarly, for type II AGN, the LDA method identifies 55 out of 214 hosts (26\% $\pm$ 3\%) with good quality data while, on the other side, 73 out of 236 hosts (31\% $\pm$ 3\%) show evidence of tidal features. Again, more than half of the major mergers show tidal features (33 coincidences), 29 belonging to S0-Sb types, 1 to Sbc type, and 3 Elliptical galaxies.

Furthermore, the configuration of the tidal features on the SDSS images could also play an important role in the identification of major mergers with the LDA method. If tidal features are present but appear as symmetric structures or do not satisfy the LDA image S/N requirements, then they may not be captured in the segmentation maps and thus will not be classified as a major merger. Therefore, it is expected, and confirmed by our results, that the galaxies classified as major mergers are not necessarily coincident with those showing tidal features or viceversa.

Among major mergers identified with the LDA method, showing at the same time evidence of tidal features, we also noticed a prevalence of early-type disks over ellipticals. This is also expected given the clear dominance of early-type disks among our mergers.

\subsection{Statistical significance of the LDA results}
\label{sec:Significance}

To test the significance of the identification of major mergers in our AGN samples, we compiled a control sample of MaNGA DR15 non-AGN galaxies by matching the morphological type, color, stellar mass and redshift distributions of our type (I+II) AGN sample. The MaNGA DR15 sample is large enough ($>$ 4500 galaxies) to permit the compilation of a control sample with $\sim$ 20 non-AGN control galaxies per matched bin. 

\begin{figure}
\centering
  \includegraphics[width=\columnwidth]{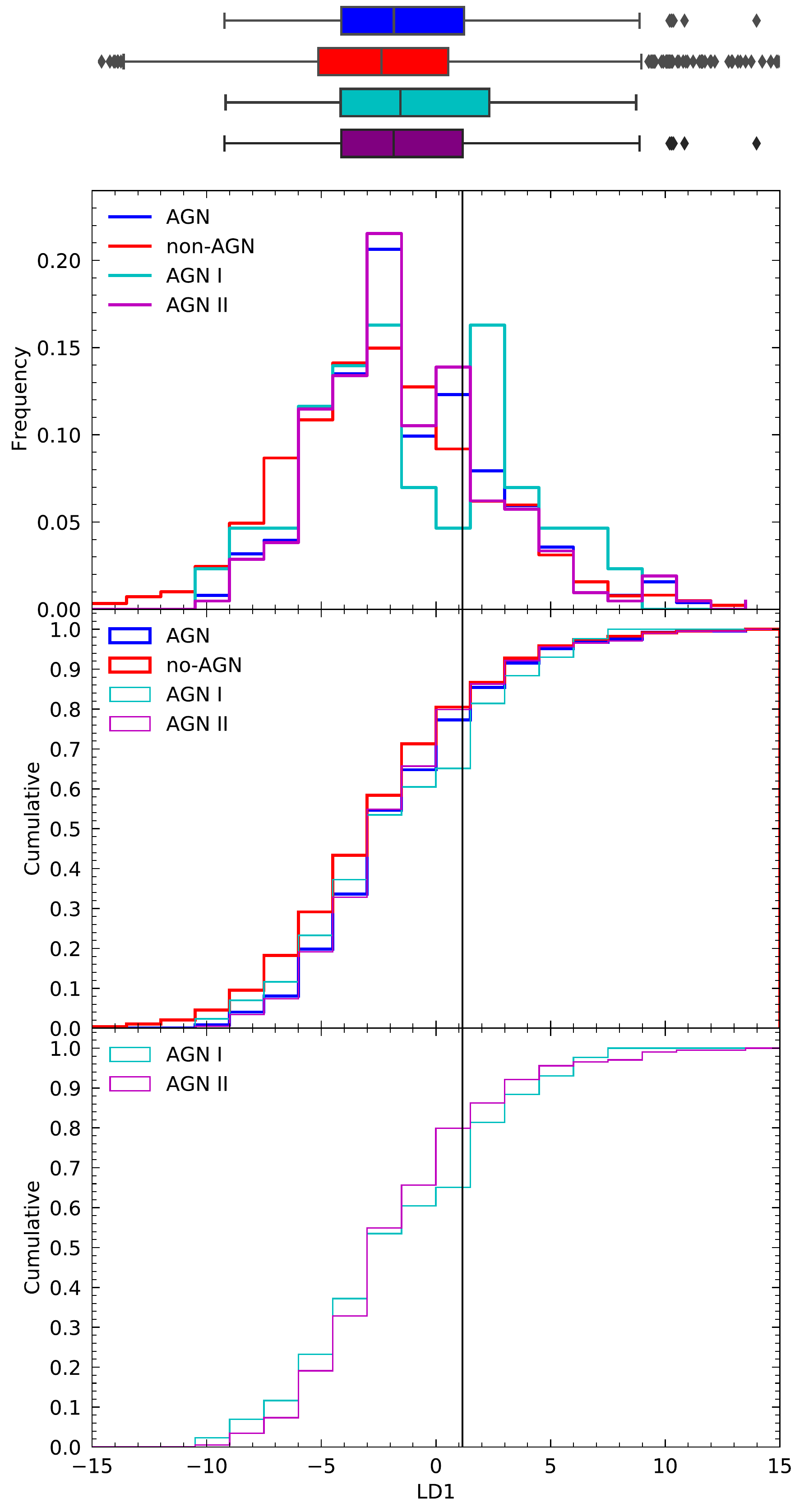}
  \caption{Cummulative LD1 parameter distributions for non-AGN (red) and AGN (blue) $p_{\rm merg}$ mass. AGN sample is subdivided by type I (cyan) and type II (purple). The top panel shown the boxplots of the LD1 distributions.
  }
\label{fig:LD1_mass}
\end{figure}

The SDSS images of galaxies in the control samples were processed in a similar way as those in our AGN samples in order to estimate the corresponding image predictors and further apply the LDA method.\footnote{The image predictor values for the galaxies in the control samples will be published as part of an update of the MaNGA morphological VAC (V\'azquez-Mata et al. 2023, in prep).} 
We find that the relative fraction of major mergers (LD1 $>1.16$) in the non-AGN control sample is 22\% $\pm$ 0.8\%, compared to 35\% $\pm$ 7\%, 26\%$\pm$ 3\%, and 29\% $\pm$ 3\% for the AGN type I, type II, and combined type (I+II) samples, respectively. The above fractions suggests a higher major merger incidence in our combined AGN sample than in non-AGN galaxies. However, it is important to assess the statistical significance of these results. 

The upper panel of Figure \ref{fig:LD1_mass} shows the resulting LD1 frequency distributions for the individual type I, type II and combined (type I+II) AGN host samples, and the control non-AGN sample. The corresponding cumulative distributions are also shown in the middle panel, while a relative comparison between type I and type II AGN hosts is attempted in the lower panel. The solid black line along the three panels is the LD1 threshold value for major merger candidates. On top of Figure \ref{fig:LD1_mass}, a set of boxplot diagrams are displayed to visualize the LD1 distribution for each sample. All values inside the whiskers belong to the LD1 distribution for each sample. The ones outside are considered outliers (grey diamonds). Only the ones with values greater than their respective U$_W$, will be considered since they are major merger candidates.

To compare the distributions we use the boxplots diagrams, a student T-test looking for differences in the mean, and the Kolmogorov-Smirnov (K-S) test. Table \ref{tab:LD1_cumu} shows the values of the different percentiles of the LD1 distributions and Table \ref{tab:test_LD1} reports the results of the K-S test after comparing the cumulative distributions and of the Student T-test for the significance of the mean.

The boxplots of the combined type (I+II) AGN and non-AGN samples show similar grouping or tightness around the corresponding median, but with a slightly more asymmetric distribution towards positive values for the combined type (I+II) AGN sample. This is reinforced by the $Q_3$ value since it is greater than the limit value (1.16) for the identification of major mergers for the combined type (I+II) AGN sample, and smaller than that limit for the control sample. Both the K-S and Student T-tests indicate that the cumulative LD1 distribution of the type (I+II) AGN sample is significantly different than that of the control sample with a mean that is slightly higher but also significant, meaning that the combined type (I+II) AGN sample shows a moderate higher (roughly of $30\%$) incidence of major mergers. 

\begin{table}
\begin{center}
\begin{tabular}{ccccc}
\hline 
\hline
& Q1 & Q2 & Q3 & IQR \\
\hline
AGN I+II&-4.13& -1.83& 1.23& 5.37\\
non-AGN&-5.13& -2.38& 0.54& 5.67\\
AGN I&-4.16& -1.55& 2.33& 6.49\\
AGN II&-4.13& -1.85& 1.16& 5.29\\
\hline 
\hline
\end{tabular}
\caption{Quartil values for the boxplots shown in Figure \ref{fig:LD1_mass}.}
\label{tab:LD1_cumu}
\end{center}
\end{table}

\begin{table}
\begin{center}
\begin{tabular}{ccc}
\hline 
\hline
sample  &  KS-test  &  T-test \\
        &  p-value  &  p-value  \\
\hline
non-AGN vs AGN I+II  &  0.01  &  0.0  \\
non-AGN vs AGN I  &  0.16  &  0.11  \\
non-AGN vs AGN II  &  0.01  &  0.01  \\
AGN I vs AGN II  &  0.21  &  0.75  \\
\hline 
\hline
\end{tabular}
\caption{KS- and T-test values for the boxplots shown in Figure \ref{fig:LD1_mass}.}
\label{tab:test_LD1}
\end{center}
\end{table}

If a comparison between type I and II AGN samples is attempted, the corresponding boxplots show that they have similar limits but different grouping or tightness around each median. The $Q_3$ values also hint that the fraction of major merger candidates (LD1 > 1.16) is greater for type I AGN compared to type II AGN. However, in this case the K-S and Student T-tests suggest non-significant differences, probably reflecting that more significant numbers in the type I AGN sample, are required for a robust comparison.

\subsection{Implications for the level of Nuclear Activity} 
\label{sec:implications}

Mergers could leave an imprint on the host galaxies by influencing the triggering of past and recent nuclear activity. Some studies support the predictions that major merging may explain the enhancement of SF activity \citep[e.g.,][]{Ellison+2008,Jogee2009,Patton2011,Scudder+2012,Ballesteros2015,Cortijo2017a,Cortijo2017b,Cortijo2017c,Thorp2019, Pan2019}, and the elevation of AGN activity \citep[e.g.,][]{Ellison2011,Treister2012,Satyapal2014,Weston2017,Hewlett2017}.  

\begin{figure*}
\centering
  \includegraphics[width=17cm]{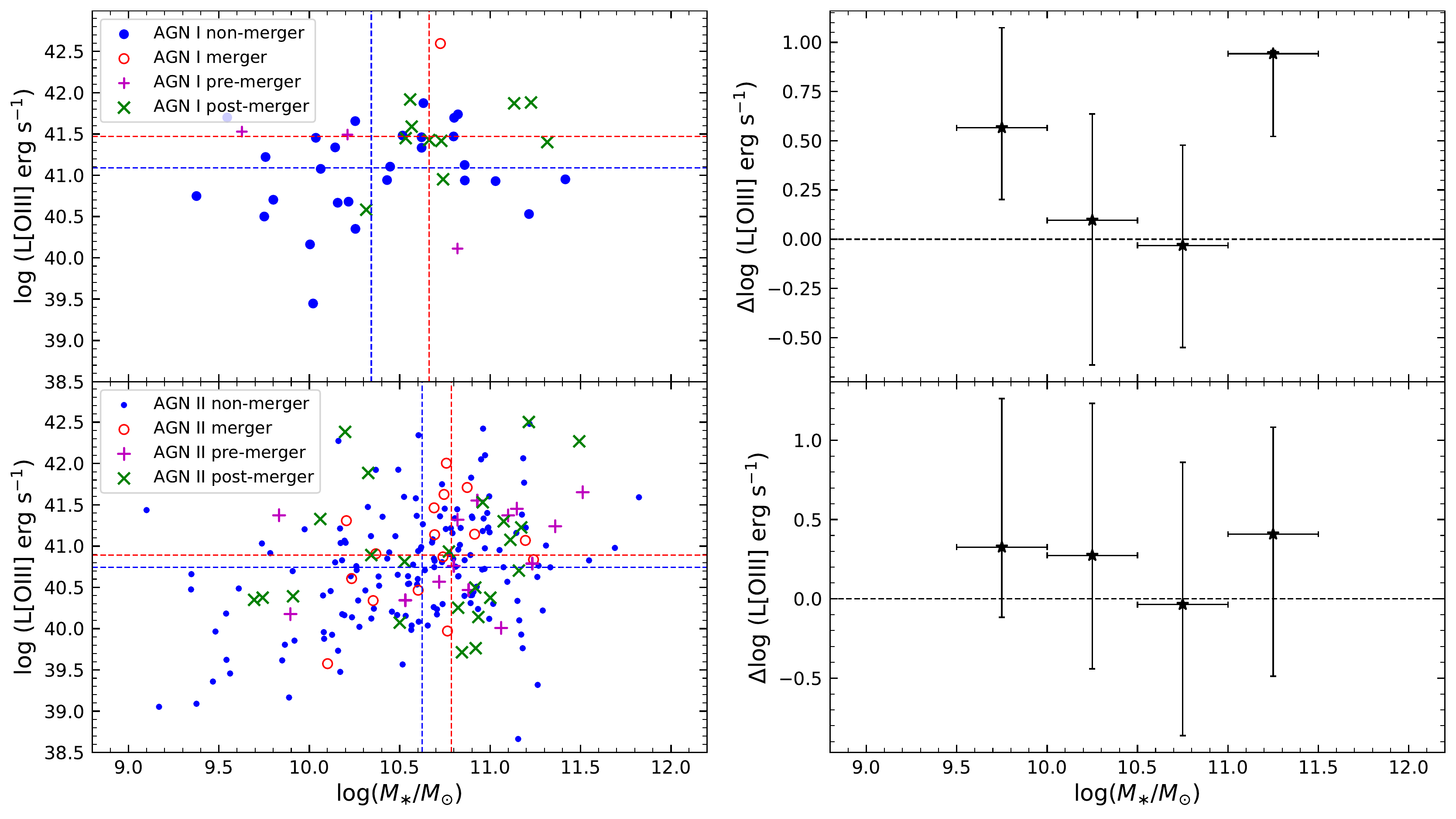}
\caption{Left-hand panels: The log(\ms/M$_{\odot}$) - log$\rm L_{ [OIII]}$  diagram for type I AGN (upper left panel) and type II AGN (lower left panel). Mergers in AGN (red symbols) as well as non-mergers in AGN (blue symbols) are highlighted along with their corresponding merger pre-coalescence (red crosses) and post-coalescence (green crosses) stage when available. The horizontal and vertical dashed lines represent the median values of log(\ms/M$_{\odot}$) and of Log$\rm L_{[OIII]}$ for mergers (red) and non-mergers (blue). The right-hand panels show the $\rm L_{[OIII]}$ enhancement,  ($\rm \Delta$ log$\rm L_{[OIII]}$ = $\rm <L_{[OIII]}>_m$ - $\rm <L_{ [OIII]}>_n$) shown by the mean of merger AGN compared to the mean of non-merger AGN, in a given 0.5 dex wide stellar mass bin.  The horizontal bars represent the bin size and the vertical bars represent the interquartil difference around the median.}
\label{fig:OIII_M_merger}
\end{figure*}

Under that scenario, we can test the expectation that SMBH in mergers could accrete material at higher rates than those found in our non-merger AGN counterparts.  \oiii\ luminosity ($\rm L_{[OIII]}$), available from our spectroscopic analysis to the MaNGA AGN samples (Cort\'es-Su\'arez et al. in preparation), can be used as a measure of the level of SMBH accretion  \citep[e.g.,][]{Heckman2014}. \oiii\ luminosities were corrected for reddening using the reddening curve from \cite{Calzetti2000}, assuming R$_V$ = 3.1 and case-B recombination, \ha/\hb = 2.86. Figure \ref{fig:OIII_M_merger} shows the log$\rm L_{ [OIII]}$-log\ms\ diagram for type I AGN (upper left panel) and type II AGN (lower left panel), highlighting in each case the identified AGN merger candidates (red) and non-merger AGN (blue) in this work. Additional symbols (red pluses for pre-mergers, and magenta crosses for post-mergers) emphasize the results of the merger stage classification from \cite{Nevin2023}. The vertical and horizontal dashed lines in each panel show the representative median log\ms\ and log$\rm L_{ [OIII]}$  values for non-mergers AGN (blue) and merger AGN (red).

The left panels in Figure \ref{fig:OIII_M_merger} show an apparent trend in type I and II AGN of increasing  log$\rm L_{[OIII]}$  values as the stellar mass increases. Mergers in AGN also appear more massive (median log\ms, red dashed vertical line) and more luminous (median log$\rm L_{[OIII]}$, red dashed horizontal line) than their corresponding non-mergers counterparts (median horizontal and vertical blue dashed lines) with higher differences shown by type I AGN (upper panel) than type II AGN (lower panel). 

The right-hand panels of Figure \ref{fig:OIII_M_merger} show the $\rm L_{[OIII]}$ enhancement  ($\rm \Delta$log$\rm L_{[OIII]}$) defined as the mean of merger AGN (log<$\rm L_{[OIII]}$>$\rm {_m}$) compared with the mean of non-merger AGN (log<$\rm L_{[OIII]}$>$\rm {_n}$) in a given mass bin,

\begin{equation}
    \rm \Delta log L_{\rm [OIII]}=log<L_{\rm [OIII]}>{_m}-log<L_{\rm [OIII]}>{_n}
\end{equation}

The horizontal error bars represent the size of the stellar mass bins ($\Delta$ Log\ms= 0.5), while the vertical error bars represent the interquartile difference (84\% - 16\%) around the median, after the $\rm L_{[OIII]}$ statistics in a mass bin.

Despite the large error bars, type II AGN (lower right-hand panel) mergers could reach a maximum enhancement of 0.4 dex (2.5$\times$) compared to non-mergers, while type I AGN mergers (upper right-hand panel) could reach a maximum about 0.9 dex ($\sim 8\times$) compared to non-merger AGN. These results are consistent with those by \cite{Ellison2013} reporting that close pairs and post-mergers show \oiii\ luminosities enhancements $\sim 3\times$ and $8\times$ higher than their corresponding control samples, respectively. However, notice that other works like \cite{Jin2021} and \cite{Steffen+2023} using pairs from the MaNGA survey find that the \oiii\ luminosity of AGN in paired galaxies is rather consistent with the AGN control galaxies (matching in mass and redshift).

The left panel of Figure \ref{fig:OIII_M_merger} also highlights the loci of the identified merger stages; pre-mergers (magenta plus symbols) and post-mergers (green crosses) in the type I AGN (upper panel) and type II AGN (lower panel) samples.
 
At this point it is important to evaluate the impact of the accuracy of the identification of major mergers in our results. If an important fraction of false mergers were included in our AGN samples, their predicted effect would be to lessen the average enhancement in \oiii\ luminosity in mergers when compared with non-mergers. \cite{Nevin2019} have tested the contamination of non-mergers in the major merger sample using their simulations finding that it is low. Specifically, the precision, which quantifies the number of true positives relative to all positives (true and false) is of the order 96\%.

\begin{figure*}
\centering
  \includegraphics[width=\textwidth]{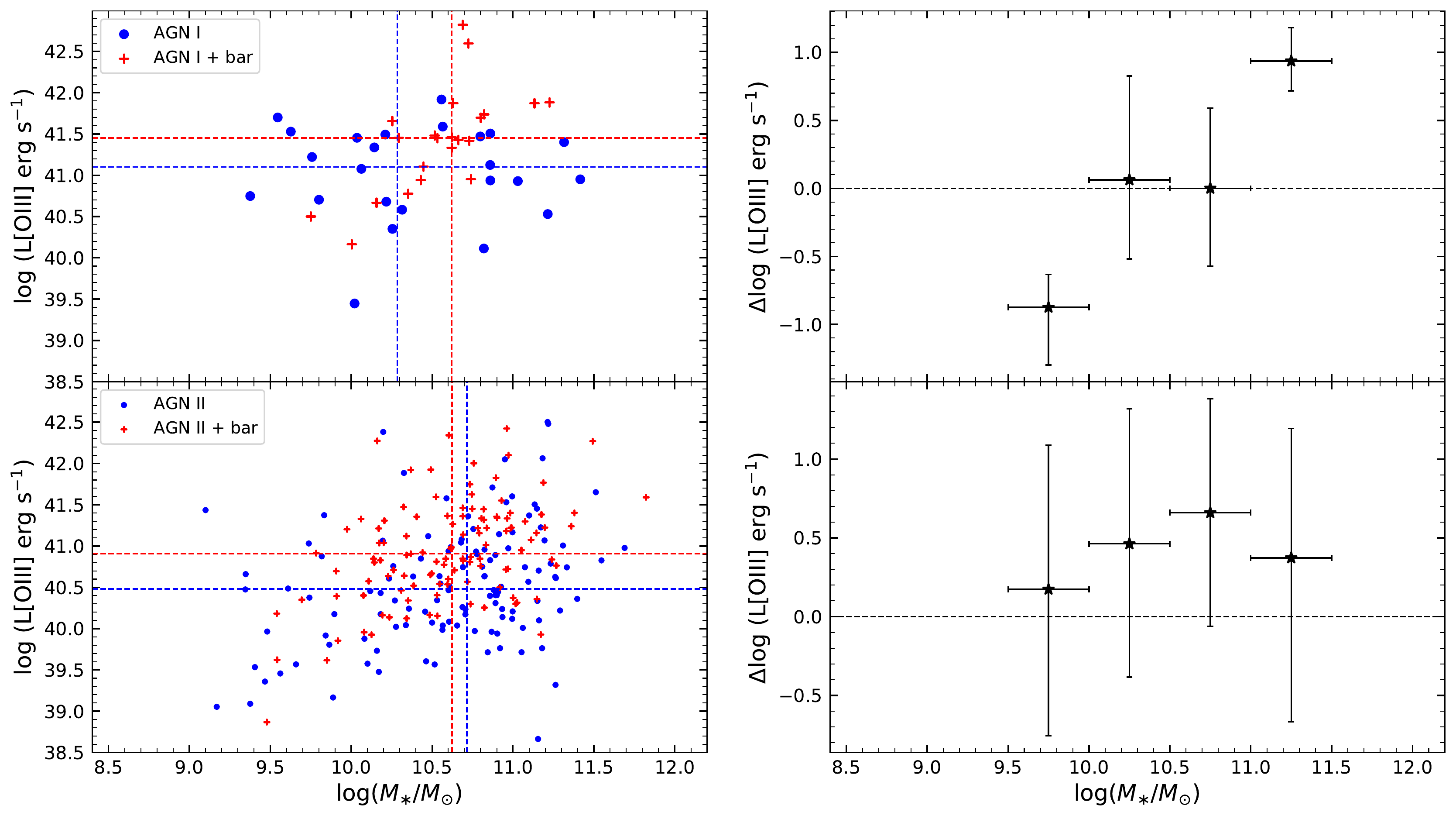}
  \caption{Left-hand panels: The log(\ms/M$_{\odot}$) - log$\rm L_{[OIII]}$ diagram for barred (red symbols) and non-barred (blue symbols) AGN hosts. The corresponding horizontal and vertical dashed lines represent the median values of Log(\ms/M$_{\odot}$) and log$\rm L_{[OIII]}$ values for barred (red) and non-barred (blue) galaxies. Right-hand panels: 
The enhancement in $\rm L_{[OIII]}$ ($\Delta$log$\rm L_{[OIII]}$=log<$\rm L_{[OIII]}$>$_b$-log<$\rm L_{[OIII]}$>$_n$) shown
by barred AGN compared to non-barred AGN. The horizontal bars represent the bin size (0.5 dex) and the vertical bars represent the interquartile difference around the median.}
\label{fig:OIII_M_bar}
\end{figure*}

\subsection{Bars and Nuclear Activity}
\label{sec:bars}

Our AGN samples show relatively lower levels of AGN activity. The median $\rm L_{ [OIII]}$ in our type II AGN is 10$^{40.74}$ erg s$^{-1}$, whereas for our type I AGN, $\rm L_{ [OIII]}= 10^{41.37}$ erg s$^{-1}$, overlapping in part but also even lower than the levels in other samples of AGNs in the local universe \citep{Oh2015,Liu2019}. If $\rm L_{ bol}=C\times \rm L_{[OIII]}$,  
where $C = 142$ for AGNs with log($\rm L_{ [OIII]}$/erg s$^{-1}$)  = 40–42  \citep{Lamastra2009}, then $L_{\rm bol}$ $\sim$ 10$^{42.90}$ and 10$^{43.52}$ erg s$^{-1}$, respectively. If we further consider representative median values \ms\ for our type I and II AGN samples from the volume-corrected mass distributions in \S\S\ \ref{sec:globalproperties} and estimate median SMBH masses through the \ms--M$_{\rm BH}$ relation \citep[e.g.,][]{Reines2015}, then our AGN samples reach Eddington ratios $\sim$ 0.005 and $\sim$ 0.030, respectively. These modest levels of AGN activity could also be sustained by internal sources of fueling \citep{Ho2009} or mediated by secular processes without external gas supply through dynamical interactions or mergers being the primary mechanism.

We look for evidence of a possible secular scenario that could explain the observed levels of AGN activity in our samples. To that purpose we use the information on the incidence of bars coming from a detailed visual morphological classification of the MaNGA sample \citep{VazquezMata2022}. While the fraction of mergers found in our combined type I+II AGN sample is about 29\% $\pm$ 3\%, more than half ($\sim$ 56\% $\pm$ 3\%) of the disk galaxies hosting our type I and II AGN samples are barred, and even the few cases of Sbc-Sc galaxies hosting our AGN are in 75\% of the cases barred. For the corresponding control sample, matched in stellar mass, color and redshift, the fraction of bars in galaxies of morphological types later than S0 is 46\% $\pm$ 2\%, similar to fractions reported in other local samples of field galaxies using quantitative bar detection methods on SDSS images \citep[e.g.,][]{Aguerri2009}. 
It is also important to evaluate the impact of the accuracy of the identification of bars in our results. Our image post-processing and the resolution of the SDSS and DESI images allowed us to detect a wide variety of bars such as bright bars with sharp ends, Ferrers and Freeman bars, and bars associated to lenses and spiral arms. A potential problem is the detection of flat bars showing a smooth transition to the disk, bars aligned with the disk, and small-sized bars embedded in prominent bulges. However, the residual images of the DESI legacy survey, the residual images after our Galfit 2D decomposition, and our filter-enhanced images, proved to be very useful for their identification. Thus, neglecting inclination effects, 
we are minimizing the presence of false positive bar detections that could affect our results. For more details on the visual identification of bars see \citet[][2023 in prep.]{VazquezMata2022}{}{}. 

Similar to Figure \ref{fig:OIII_M_merger}, the left-hand panels in Figure \ref{fig:OIII_M_bar} highlight barred AGN (red symbols) and non-barred AGN (blue symbols). The vertical and horizontal red and blue dashed lines represent their corresponding median log(\ms/M$_{\odot}$) and log$\rm L_{ [OIII]}$ values. 
The left-hand panels show an apparent global trend of the barred type I and II AGN to have higher log$\rm L_{ [OIII]}$ values as the stellar mass increases, with Pearson correlation coefficients $p = 0.72$ and $p = 0.46$ for type I AGN and type II AGN, respectively.

By defining the level $\rm L_{[OIII]}$ enhancement in a similar way as that in Figure \ref{fig:OIII_M_merger}, an internal comparison between barred and non-barred type I AGN shows that 
bars in type I AGN produce a significant luminosity enhancement when compared to non-barred type I AGN (Student t-test p = 0.03). Similarly, bars in type II AGN also produce a noticeable luminosity enhancement when compared to non-barred type II AGN (Student t-test p = 2 $\times$ 10{-5}). 

A comparison of the effects produced by mergers and bars in Figures \ref{fig:OIII_M_merger} and \ref{fig:OIII_M_bar} shows that type II barred AGN (lower right-hand panel in Figure \ref{fig:OIII_M_bar}) reach a maximum level of $\sim$ 0.7 dex ($5\times$) compared to non-barred AGN, higher than the maximun level ($2.5\times$) found in type II AGN mergers, whereas in type I AGN (upper right-hand panel in Figure \ref{fig:OIII_M_bar}) the bar enhancement reach a maximum of about 0.9 dex ($\sim 8\times$), comparable to the maximum level found in type I AGN mergers. 

These results indicate that the AGN activity in our low luminosity AGN samples is not uniquely promoted by merger events but could be contributed by bars. Since we have evaluated that the contamination of non-mergers and false bars is not important and thus, is not impacting the observed levels of enhancement, we conclude that bars are also playing a fundamental role in the AGN stimulation among our AGN samples. In line with our results, \cite{Alonso2018} have shown that the dynamical perturbations produced by interactions and mergers (previous to full coalescence) and bars produce an enhancement in nuclear activity and accretion rate in AGN galaxies with bars being a more efficient mechanism than interactions. Furthermore, they report that the efficiency of the mergers and interactions in transporting material towards the inner regions of galaxies depends not only on the properties of the hosts but also strongly on the perturber companion properties. When the perturber companion tends to be massive, luminous, and with high gas content, the effect of mergers and interactions on the central nuclear activity tends to be as efficient as that induced by bars. Thus, a more appropriate comparison of the levels of enhancement produced by bars and mergers (those in stages previous to full coalescence) should take into account the properties of the perturber companions. However, that is out of the scope of the present paper. Another important factor to take into account in this comparison is the difference in timescales of the processes involved, something debated in section \ref{sec:discussion}.

If we look for the fraction of major merger AGN that are simultaneously barred, we find 53\% $\pm$ 13\% for type I AGN and 55\% $\pm$ 7\% for type II AGN. On the contrary, if we look for the fraction of barred AGN that simultaneously are major mergers we find 35\% $\pm$ 10\% for type I AGN and 23\% $\pm$ 4\% for type II AGN. Thus, within our AGN samples, being a merger and simultaneously having a bar is more frequent than having a bar and simultaneously being a merger.

\cite{Peirani2009} and \cite{Moetazedian2017} have shown that merging galaxies can induce bars prior to the collision, however, very few is known about how bars can emerge from the aftermath of a merger. \cite{Cavanagh2020} considered three generic ways of bar formation: (i) the spontaneous self-gravitating isolated model, (ii) the tidal interaction model, and (iii) the galaxy merger model with two bar formation phases identified during galaxy merging. They find that mergers with low mass ratios and closely-aligned orientations are considerably more conducive to bar formation compared to equal-mass mergers. They also find that it is possible for a bar to regenerate in the case of nearly equal spin angles in a major merger, inferring that the transfer of angular momentum is key to the regeneration of the bar.

Bar formation appears as a process naturally linked to the merger process so that it is expected that an important fraction of the major mergers (either in pre- or post-coalescence stage) in this work show at the same time a bar, as shown by our results. On the other side, by looking at the fraction of barred AGN that simultaneously are major mergers, since the bar formation process is not only related to major mergers, it is thus expected that a lower fraction of bars in our AGN samples are associated only to major mergers, as also shown by our results. Another possibility is that the LDA method is somewhat biased to assume bars are in non-interacting galaxies or minor mergers. Although \cite{Nevin2023} does not address the existence of a bias with bars in the LDA method, we discard such a bias since the LDA method relies on morphology with training sets including weak and strong bars in both the merger and non-merger samples, thus making it not biased against bars.

\subsection{The Local and Large Scale Environment}

Finally, we investigate potential differences that may be associated to the environment and that could be affecting the results of a statistically significant higher incidence of major mergers in our AGN samples compared to a control non-AGN sample. Galaxies are also exposed to the influences of their local and large-scale environment. Galaxy properties correlate with the Large-Scale background density Structure (LSS) at low redshift \citep[e.g.,][and reference therein.]{Park2009,Muldrew2012}, and there is also evidence linking the presence of AGNs to the local environment \citep[e.g.,][and references therein]{Ellison2013}. On the other hand, other studies \citep{Sabater2013,Sabater2015} suggest that large-scale environment and galaxy interactions play a fundamental but indirect role in AGN activity (by influencing the gas supply). 

\begin{figure*}
\centering
  \includegraphics[width=\textwidth]{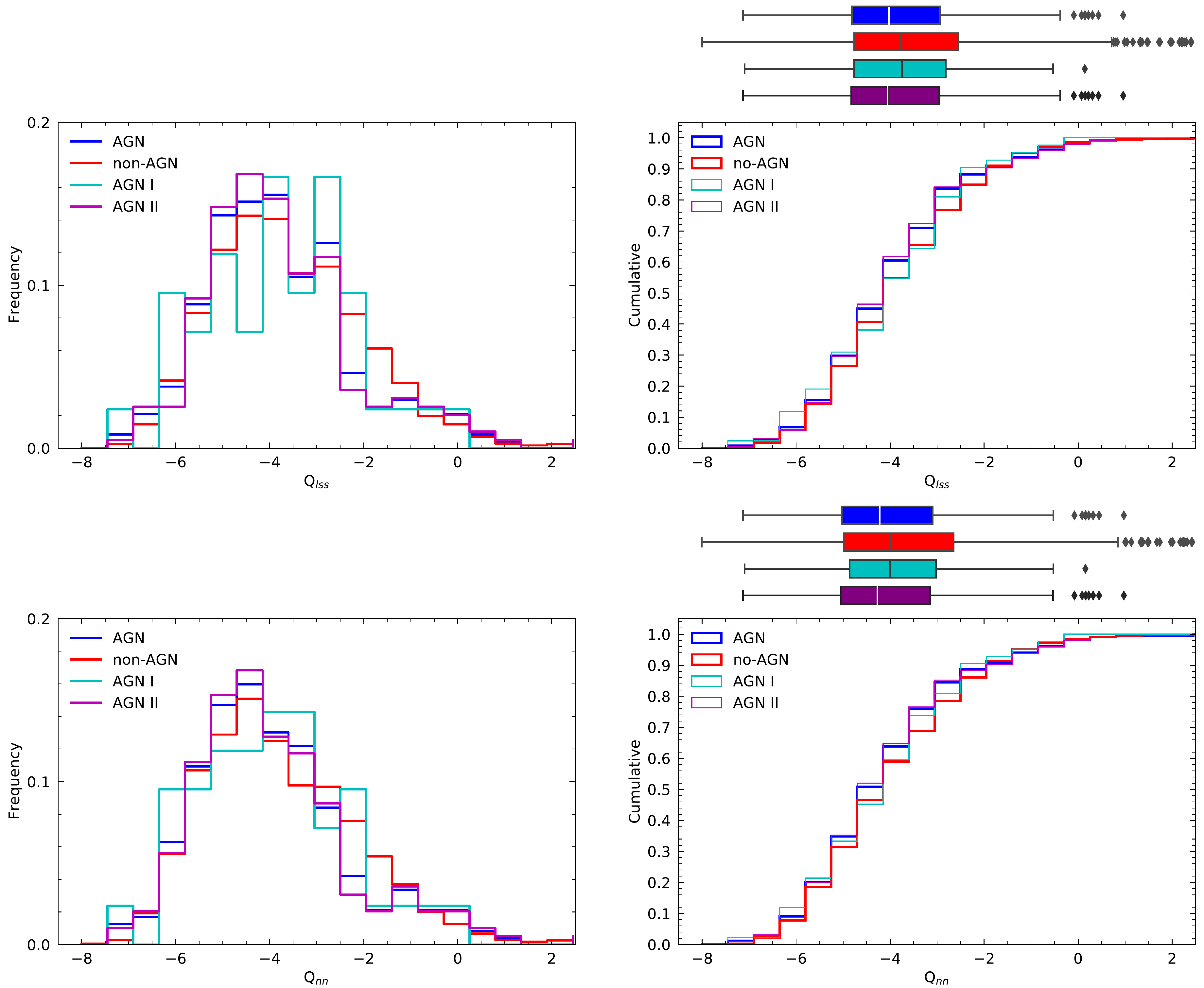}
  \caption{Left-hand panels: The frequency distributions of the environment indicators $Q_{\rm lss}$ (upper panels) and $Q_{\rm nn}$ (lower panels) for our type I (cyan), type II (purple), combined (type I+II; in blue) and non-AGN control MaNGA DR15 sample (red). Right-hand panels: The  cumulative distributions of the same indicators. On top of the right columns the corresponding boxplots diagrams are also shown.  
  }
\label{fig:1Mpc-Environment}
\end{figure*}

To this purpose we compiled a non-AGN control sample, by matching only the redshift interval in order of not biasing its environment properties. To proceed, we adopted various environment quantifications reported in the Galaxy Environment for MaNGA Value Added Catalogue (GEMA-VAC; Argudo-Fernández et al., in prep.) based on the methods described in \cite{Argudo2015}. 

We consider the information on the tidal strength parameter $Q$, defined as an estimation of the total gravitational interaction strength that the neighbours produce on a target central galaxy with respect to its internal binding forces. We also consider those estimates on various scales namely: (i) the $Q_{\rm nn}$ parameter defined as the tidal strength of the 1st nearest neighbour and, (ii) a more global characterization of the Large Scale Structure (LSS) environment by all the neighbours within 500 km/s line-of-sight velocity difference, up to 5 Mpc projected distances in a volume limited sample up to z < 0.15 traced by the parameter $Q_{\rm lss}$, defined as the tidal strength of the LSS. Notice that the environment probed with these parameters do not include the influence of much more denser regions, where differences could be found. 

Figure \ref{fig:1Mpc-Environment} shows on the left column the frequency distributions and on the right column the cumulative distributions of the environment indicators $Q_{ \rm lss}$ (upper panels) and $Q_{ \rm nn}$ (lower panels)  for our type I (cyan), type II (purple), combined (I+II; in blue) and non-AGN control MaNGA DR15 sample (red). On top of the right columns the corresponding boxplots diagrams are also shown. Tables \ref{tab:quartil_env} and \ref{tab:testEnv} summarizes the percentiles of the distribution for each environmental parameter and the results of a K-S and Student t-test after comparing the corresponding cumulative distributions and the significance of the mean. 

For the interpretation of the frequencies and cumulative distributions, we adopt representative values for $Q_{ \rm lss}$ for isolated galaxies, isolated pairs and isolated triplets reported in \cite{Argudo2015}. We also built frequency distributions of galaxies in a range of environments from poor groups ($3 < n < 15$), rich groups ($15 < n < 50$) and clusters ($n > 50$) catalogued in \cite{Yang2007} and in common with the MaNGA sample in order to find representative $Q_{ \rm nn}$ and $Q_{ \rm group}$ values for these environments.  

According to \cite{Argudo2015} the representative environment $Q_{ \rm lss}$ values for isolated pairs and isolated triplets range from $-5.5$ to $-5.0$, with values smaller than $-5.5$ associated to galaxies in more isolated environments, and values greater than $-5.0$ associated to poor groups and increasingly rich groups. According to this, the representative $Q_{ \rm lss}$ environment of our type I, II, and non-AGN samples goes from that of poor groups to that of intermediate/rich groups, and our comparison in Tables \ref{tab:quartil_env} and \ref{tab:testEnv} shows that they are non-significantly different. The percentile values show slight differences between samples in both $Q_{ \rm lss}$ and Q$_{\rm nn}$. However, for the $Q_{ \rm lss}$ distribution we can see that lower percentile values, Q$_1$, are higher than $-5$, which means that less than 25\% of the combined type (I+II) AGN and control samples are in isolated environments. 

Representative $Q_{ \rm nn}$ values are slightly lower than $Q_{ \rm lss}$ values since for $Q_{ \rm nn}$ only the nearest galaxy is considered, exerting typically of the order of 90\% of the total tidal force. Thus, in the lower panel a slight shift to lower values compared to $Q_{ \rm lss}$ are observed corresponding to poor and intermediate/rich groups. Our comparisons also show non-significant differences in the cumulative and mean of type I, II, and non-AGN samples.

Notice that parameters like $Q_{ \rm group}$ although useful, were not considered here due to a lack of available information. According to our tests, we cannot reject the null hypothesis that our type I, II, combinned type (I+II) and non-AGN samples are drawn from the same population, with non-significant differences in the mean, concluding that these samples share a similar local and LSS environment, discarding possible influences of the local environment in our results on the incidence of mergers in our AGN samples. A more detailed study of the Large Scale Environment of MaNGA galaxies by using a geometric characterization of the cosmic web with methods described in \cite{AragonCalvo2007} is reserved for a forthcoming analysis.

\begin{table}
\begin{center}
\begin{tabular}{ccccc}
\hline 
\hline
& Q1 & Q2 & Q3 & IQR \\
\hline
\hline
\multicolumn{5}{|c|}{$Q_{\rm lss}$} \\
\hline
AGN I+II&-4.81 & -4.03 & -2.94 & 1.88\\
non-AGN&-4.77 & -3.78 & -2.56 & 2.21\\
AGN I&-4.76 & -3.75 & -2.85 & 1.90\\
AGN II&-4.83 & -4.04 & -2.95 & 1.88\\

\hline
\multicolumn{5}{|c|}{$Q_{\rm nn}$} \\
\hline
AGN I+II&-5.02 & -4.23 & -3.10 & 1.92\\
non-AGN&-4.99 & -4.00 & -2.66 & 2.34\\
AGN I&-4.83 & -4.00 & -3.05 & 1.78\\
AGN II&-5.06 & -4.27 & -3.15 & 1.91\\

\hline 
\hline
\end{tabular}
\caption{Quartil values for the boxplots shown in Figure \ref{fig:1Mpc-Environment}}
\label{tab:quartil_env}
\end{center}
\end{table}

\begin{table}
\begin{center}
\begin{tabular}{ccc}
\hline 
\hline
sample  &  KS-test  &  T-test \\
        &  p-value  &  p-value  \\
\hline
\multicolumn{3}{|c|}{$Q_{\rm lss}$} \\
\hline
non-AGN vs AGN I+II  &  0.11  &  0.12  \\
non-AGN vs AGN I  &  0.83  &  0.51  \\
non-AGN vs AGN II  &  0.16  &  0.16  \\
AGN I vs AGN II  &  0.53  &  0.99  \\
\hline
\multicolumn{3}{|c|}{$Q_{\rm nn}$} \\
\hline
non-AGN vs AGN I+II  &  0.06  &  0.14  \\
non-AGN vs AGN I  &  0.78  &  0.57  \\
non-AGN vs AGN II  &  0.1  &  0.17  \\
AGN I vs AGN II  &  0.83  &  0.94  \\
\hline 
\hline
\end{tabular}
\caption{KS- and T-test values for histograms shown in Figure \ref{fig:1Mpc-Environment}.}
\label{tab:testEnv}
\end{center}
\end{table}

\section{Discussion}
\label{sec:discussion}

Previous statistical studies on the AGN-merger connection have found conflicting results as reviewed in \citet[][]{Ellison2019} and \citet[][]{Gao2020}. However, as these authors discuss, it is important to consider that these types of studies have used different sample selection and AGN criteria, as well as different methods for identifying mergers. The above introduces various and different selection biases in the results of each study. 
First of all, we emphasize that depending on the observational sample and the method used for the study, different stages of the merging process are considered, from early pre-merger stages ($< 0.5-1$ Gyr before the coalescence) to late post-merger stages (up to $\sim 2$ Gyr after the coalescence). On the other hand, the timescales of the AGN phenomenon are usually short, no more than 0.2-0.3 Gyr, and more commonly of 0.01-0.1 Gyr \citep[][]{Marconi2004,Hopkins2009}. A relevant question then is at what stage of the merger process is the possible triggering of the AGN most likely; this is something that depends on the parameters of the interaction and on the gas content of the interacting/merged galaxies, to mention some factors. As discussed in \cite{McElroy+2022} and \citet[][and more references therein]{Chang+2022}, and related also to the merger stages in which AGN are most likely, the diversity of results on the AGN-merger connection may be associated to: (i) the galaxy samples used for the study (galaxy pairs in spectroscopic redshift surveys, optical images of morphologically disturbed galaxies, etc.), (ii) the diversity of predictors of interaction/merger, (iii) the type of observations (photometric, kinematic, etc.), and (iv) the type of merger (minor, major, other).  Also, as discussed and shown in \citet[][]{Ellison2019} and \citet[][see also \citealp{Ji2022}]{Gao2020}, the selection criteria for AGN (e.g., mid IR color, X-ray, optical emission line ratios, radio) affects the results of studies on the AGN-merger connection, probably because the different AGN selection criteria may represent different merger stages \citep[e.g.,][]{Sanders1988a} and different types of AGN, especially different AGN luminosities.

In the present paper, we have studied the incidence of \textit{major mergers} in an {\it optically-selected} AGN sample by using the automatic LDA method, which is based on quantifying \textit{ morpho-structural distortions in optical images}.  This method allows us to identify a wide range of merging stages, from pre- to post-coalescence (see \S\S\ \ref{sec:merging-stages}), which can increase the incidence of identified major mergers in AGN hosts relative to other approaches sensitive only to a given merging stage. As shown in \S\S\ \ref{sec:major-mergers}, a major fraction  of those AGN host galaxies identified as major mergers are in the post-coalescence stage (a factor of $\sim 2$ more than in the pre-coalescence stage), which suggests that the triggering of AGNs is more probable after the coalescence phase \citep[see also][]{Ellison2013}. 
From an approach in the opposite direction, various studies of galaxy mergers at low redshift find a statistically significant enhanced AGN fractions. \citet{Carpineti2012} explored a subset of post-mergers, where a single remnant is in the final stages of relaxation after the merger, finding an important rising of the AGN fraction and suggesting that the AGN phase probably becomes dominant only in the very final stages of the merger process. \cite{Bickley2023} quantified the frequency of AGN in fully coalesced post-merger systems, and further compared this frequency to that in a sample of galaxy pairs. They find that AGN identified by narrow-line optical emission and mid-IR colour have an incidence rate in post-mergers with excesses of $\sim$4 over control samples, also exceeding the values found for galaxy pairs, indicating that AGN activity in mergers peaks after coalescence. Furthermore, the \oiii\ luminosity in post-mergers that host an optical AGN is $\sim$0.3 dex higher on average than in non-interacting galaxies with an optical AGN, suggesting that mergers generate higher accretion rates than secular triggering mechanisms.

The identification of \textit{major merger} morphological signatures in the present work is more sophisticated and detailed than some previous attempts using similar optical imaging. 
We have identified a fraction of major mergers going from 25\% to 35\% in the type II and type I AGN samples, respectively, with a value of 29\% $\pm$ 3\% for the combined type (I+II) sample, compared to the value of 22\% $\pm$ 0.8\% for the non-AGN control sample. This result shows evidence of a higher incidence of major mergers in AGN galaxies than in non-AGN ones, supporting the idea that an external mechanism through galaxy merging can trigger or enhance the AGN activity in low-$z$ galaxies, but this mechanism appears not to be dominant.

\subsection{Comparison with previous works}
\label{sec:comparisons}

The higher incidence of major mergers found in our AGN sample with respect to the control non-AGN sample seems to be more significant than those reported in other studies with similar stellar masses, morphological content, and image quality \citep{Darg2010,Ellison2011,Ellison2013,Satyapal2014,Weston2017,Mantha2018,Thibert2021}.
However, note that these studies looked for the incidence of AGNs in merger galaxy samples, while we looked for the incidence of mergers in AGN (and non-AGN control) samples. As discussed in \citet[][]{Ellison2019}, the former studies are focused on exploring whether mergers can trigger AGN or not, providing an affirmative answer to this question at a statistical level, while the later studies focus on exploring the degree of importance of mergers on triggering AGNs. On the other hand, most of these works (i) used close galaxy pairs to select interacting galaxies, such that the incidence refers to the pre-coalescence and (early) post-coalescence phases, and (ii) identified merging features with only a few image predictors. 

As mentioned above, the LDA method is able to detect a wide range of merging stages, from  pre- to post-coalescence, dominating the latter. The method is more sensitive to identify merger signatures than these previous works with major merger observability timescales $>2Gyr$ since it incorporates the strengths of various image predictors, and it takes into account implicitly different gas fractions. Related to the later, \cite{Lotz2011} found that asymmetry was detectable in timescales of $<$ 300 Myr for f$_{\rm gas}$ $\sim$ 20\%  increasing to about 1 Gyr for f$_{\rm gas}$ $\sim$ 50\%. In such circumstances, gas-poor galaxies may have had recent mergers or interactions $<$ 300 Myr ago but appear not identifiable as major mergers if only one image predictor is used.

The studies by \citet[][]{Ellison2019} and \citet[][]{Gao2020} are more similar to ours in that (i) they look for merger incidence in selected samples of AGN galaxies, and (ii) to identify mergers in their AGN and non-AGN samples they evaluate morphological distortions in optical images. \citet[][]{Gao2020} use the SDSS DR7 \citep{Abazajian2009} and GAMA \citep[][]{Liske2015} surveys, both optical BPT diagram and mid-IR color criteria to select AGNs, and a deep learning convolutional neural network (CNN) technique trained with visually identified merging galaxies within Galaxy Zoo \citep[][]{Lintott2008}, GAMA-KiDS Galaxy Zoo \citep[][]{Holwerda2019}, and also using the smoothness and asymmetry statistics \citep[][]{Conselice2003}.
\citet[][]{Gao2020} find that about 16\% of the optical AGN hosts in the SDSS sample show evidence of mergers versus a fraction of 14\% for the control non-AGN sample. If the GAMA survey is used instead, 39\% of the optical AGN hosts show evidence of mergers versus a fraction of 28\% for the control non-AGN sample. When they use the mid-IR criteria to select AGN, they find that about 23\% of their host galaxies in the SDSS sample show evidence of mergers versus a fraction of 15\% for the control non-AGN sample. While if the GAMA survey is used, 39\% of the mid-IR AGN host galaxies show evidence of mergers versus a fraction of 28\% for the control non-AGN sample. The differences between the SDSS and GAMA samples could be due to the deeper imaging of KiDS revealing subtle features, the higher redshift range in the GAMA sample, and/or differences in the training sample used in the CNN for each survey. The above shows a non-negligible to potentially moderate role of mergers in triggering AGNs, with the merger fraction increasing as stellar mass increases. 

\citet[][]{Ellison2019} identify AGN in a SDSS sample also using optical emission line diagnostics and mid-IR colours. For the merger characterization, they look for signs of morphological disturbance such as tails or shells and/or the presence of a perturbing companion. Those galaxies lacking any obvious morphological disturbance and without companions are classified as isolated. The control non-AGN sample is gathered with galaxies that are matched in \ms\ and $z$. \citet[][]{Ellison2019} find that $37\%$ of optically selected AGN host galaxies show signs of visual disturbances or have close companions.\footnote{Compare this fraction to our 29\% fraction using LDA or 31\% using identification of tidal features in the SDSS images.}  In contrast, for the hosts of mid-IR selected AGNs, this fraction is $61\%$. Both mid-IR and optically selected AGN have interacting fractions that are a factor of two greater than the corresponding non-AGN control samples. 
 
Summarizing, the results by \citet[][]{Gao2020} (CNN method) for their GAMA sample and by \citet[][]{Ellison2019} (visual evaluation) for their SDSS sample, show excess of merger galaxies in optical AGN hosts with respect to non-AGN galaxies, but the incidence of mergers on the AGN host samples are moderate as to conclude that they are the dominant mechanism of optically-selected AGN triggering, in qualitative agreement with our results. However, for the samples of obscured-AGN hosts, the incidence of mergers and close companions is more relevant, showing that obscured AGNs might be  preferentially associated with merging triggering.

Other studies find higher fractions of mergers among their AGN samples, showing the importance of the image quality. \cite{Hong2015} used deep optical images from various telescopes finding that 17 of 39 luminous ($M_R$ < -22.6 mag) type I AGN host galaxies (43.6\%, compared with the fraction of $\approx 35\%$ found here) show evidence for current or past mergers in the form of tidal features and disturbed morphology. They find that the merging fraction could be even higher, after correcting for redshift effects, suggesting that for luminous AGNs, there is a much more likely association with mergers.

So far, we have shown that our AGN samples (although with moderate but still significant levels) are more frequently associated with mergers than the non-AGN control sample, suggesting a merger-AGN connection for our local samples. Furthermore, we have found that among the mergers identified in our AGN hosts, the more massive ones could trigger moderate levels of AGN activity (measured by \oiii\ luminosities) up to 0.4 dex (2.5$\times$) in type II AGN compared to non-merger AGNs, and up to 0.9 dex ($\sim 8\times$) in type I AGN  (see \S\S\ \ref{sec:implications}. These levels are comparable to those reported in \cite{Ellison2013} for paired galaxies at small separations and post-merger galaxies, when compared to their corresponding control samples.

\subsection{What is the dominant mechanism of AGN triggering?}

The results presented here, using the LDA method, show an statistical excess of $\approx 29\%$ of major merger signatures in our optical type (I+II) AGN sample with respect to the control non-AGN sample. This moderate excess shows that {\it major mergers can trigger AGN but that they are not the dominant mechanism.}  Furthermore, we have estimated statistically that the major incidence of mergers in the AGN hosts is in the advanced (post-merger) phases (51\% $\pm$ 6\% of the cases). 
For our visual identification of bright tidal features, the fraction of AGN hosts with these signatures when using SDSS images is $\approx 30\%$ ($\approx 35\%$ when using DESI images), a factor of two higher than for the non-AGN galaxies. Once again, these statistical results confirm that the AGN phenomenon can be partially related to the merging activity but this does not appear to be the main channel of AGN triggering.  

However, we were able to find some evidence of increased AGN power  
among our AGN hosts, but particularly in the most massive ones with signatures of major mergers compared to those hosts without such signatures. The above is probably partially related to the different timescales involved in these processes. On one side, observations and simulations have shown that merging events drive important amounts of gas towards the central regions of galaxies causing a starburst and feeding the AGN with typical lifetimes up to 200–300 Myr  \citep{Wild2010,DiMatteo2005,Schawinski2015}. 
Furthermore, detailed numerical simulations \citep{Hopkins2005}, show that the visibility timescales of the black hole activity could be even shorter. They predict a buried phase during the starburst, where the black hole is heavily obscured by the surrounding gas and dust, limiting its visibility at optical and ultraviolet wavelengths, showing that between the buried and ending phases of black hole activity, a galaxy would be seen as a luminous quasar, with short observable lifetimes, depending on waveband and luminosity threshold; typically 10 Myr for bright quasars in the optical $B$ band. 
On the other side, the observability window for the LDA method translates into an identification of mergers in timescales of $>$ 2 Gyr. Thus, the differences in timescales involved suggest that only a fraction of those AGN hosts with merger signatures could show evidence of enhanced levels of AGN activity at the current observation \citep[e.g.,][]{Villforth2014,Shabala2017}, explaining the trends and scatter observed in both panels of Figure \ref{fig:OIII_M_merger}. 

Our results show relatively modest probabilities that AGN triggering is due to major mergers. What about the contribution to AGN triggering by secular processes associated to bars? \citet[][]{Treister2012} have found that the most luminous AGN phases are connected to major mergers, while the less luminous AGNs, appear to be driven by secular processes. We have found that $\approx 56\%$ of our type (I+II) AGN hosts with disks are barred, a fraction significantly larger than their major merger incidence of 29\% $\pm$ 3\%.  Therefore, at least statistically, bars appear to be as a relevant mechanism of AGN triggering as major mergers. Notice, however, that for the corresponding control non-AGN sample, the bar fraction, $\approx 45\%$, is not too different to that of the AGN sample.    

\cite{Ellison2016} have shown that while secular processes predominantly lead to moderate accretion rates and are not accompanied by an increase in SFR, galaxy-galaxy interactions lead to an increase in SFR, with more powerful and possibly obscured AGN. \cite{Alonso2018} used a SDSS sample to study the influence of strong bars on AGN, also comparing the effects of interactions on activity, finding that bars and interactions increase the AGN luminosity and accretion rate, but with bars having a greater efficiency in the process.  

If bars are a relevant AGN triggering mechanism, why there so many barred galaxies with unobserved AGNs? As discussed in \cite{Alonso2018}, the presence of a bar is not enough. Gas needs to be available to be funneled to the central regions where the presence of inner structures and dynamical resonances may also be of importance for this process. Whether AGN could be a recurrent phenomena, the funnelling of gas to the centre of the galaxy ($10^{8}$ yr) and the lifetime of bars ($\sim 10^{10}$ yr) are expected to be much longer than the timescale for AGN activity ($10^{7}$ yr) imposing severe restrictions for their occurrence.

\section{Conclusions}
\label{sec:conclusions}

We have considered an optically-selected sample of 47 type I and 236 type II AGN from the MaNGA DR15 \citep{Cortes2022} at redshifts $z < 0.15$ and took advantage of the Linear Discriminant Method  presented in \citet[][]{Nevin2019,Nevin2023} to identify major mergers and merger stages using the SDSS images. To reinforce our analysis we have used detailed morphological information coming from our post-processing to the SDSS and DESI Legacy images including the identification of bright tidal features \citep{VazquezMata2022}. Along our study we built various control samples to study and compare global physical, morphological and environmental properties and the incidence of major mergers in our AGN samples. Major mergers were found as important promoters of the AGN activity, however, evidence is found that stellar bars are also playing an important role in the triggering of our AGN samples.

Our main results and conclusions are as follows.

$\bullet$ For the type I and II AGN hosts and the control non-AGN galaxies,  we reported volume-corrected morphological types, stellar masses, and ($g -i$) colors. The AGN hosts are mostly early type discs (Sa, Sb and very few Sc) with a small fraction of elliptical galaxies (12\% and 13\%, for type I and II AGN hosts respectively) but no types later than Sc. They typically inhabit earlier Sa types, compared to Sb types for non-AGN control galaxies. Their mass and color distributions are significantly concentrated toward more massive and redder colors than those of the non-AGN control sample. 

$\bullet$ The LDA method has identified a fraction of major mergers going from 25\% to 35\% in our type II and type I AGN samples, respectively, with a value of 29\% $\pm$ 3\% for the combined type (I+II) sample. The identification of major merger morphological signatures in the present work is more sophisticated and detailed than some previous attempts using similar optical imaging.
There is a modest but statistically significant higher fraction of major mergers in our type (I+II) AGN sample, 29\% $\pm$ 3\%, compared to 22\% $\pm$ 0.8\% for the non-AGN control sample (matched in stellar mass, morphology, color and redshift), supporting the idea that an external mechanism through galaxy merging can induce the observed AGN activity,  but it is not the dominant mechanism. Following \cite{Nevin2023}, we also have found a prevalence of post-coalescence (51\% $\pm$ 6\%) over pre-coalescence (26\% $\pm$ 5\%) stages in our identified major merger host galaxies.

$\bullet$ From our visual identification of bright ($<25$ mag arcsec$^{-2}$) tidal features in the AGN samples using the $r$-band SDSS images, we have found a fraction of 31\% $\pm$ 3\% for the combined type (I+II) AGN sample. This is a factor of two higher incidence compared to that in the non-AGN control sample. Since the LDA method was not trained on the basis of tidal features, we confirmed that not all major mergers show evidence of tidal features nor all galaxies having tidal features were identified as major mergers.

We also investigate other triggering mechanisms for the AGN.

$\bullet$ More than half of our AGN hosts are barred (56\% $\pm$ 3\%) for type II, and 56\% $\pm$ 8\% for type I) in contrast to a fraction of 46\% $\pm$ 1\% for the control sample.  Even the few Sbc-Sc hosts are 75\% barred. Therefore, bars can also be a (internal) mechanism of AGN trigger. 

$\bullet$ The more massive type II major merger AGN hosts show an \oiii\ luminosity enhancement up to 0.4 dex (2.5x) compared to the more massive type II non-merger AGN hosts, reaching about 0.9 dex (8x) in the more massive type I merger AGN, compared to the more massive type I non-merger AGN, these levels comparable to those reported in other works.

$\bullet$ The more massive type I and II AGN hosting stellar bars reach \oiii\ enhancement levels comparable to those found in our identified major merger AGN, suggesting that the observed AGN activity could also be sustained by internal processes, promoted by bars. 

$\bullet$  No significant differences were found between the AGN and non-AGN samples  for the tidal strength parameter $Q$, estimated at group and on large ($5$ Mpc) scales. This suggests that our results on the incidence of major mergers on AGN hosts are not biased by the local and large-scale environment. 

$\bullet$ A classification of the appearance of the central optical spectra in type I and II AGN hosts through the $H-H\beta$ indices can be used as an empirical predictor of the amount of contaminant flux coming from a central AGN source (see Appendix \ref{appendix:A}) measured in optical images. This potential correlation could be useful to first order correct a series of photometric quantities like colors, absolute magnitudes and color-dependent stellar masses in large AGN samples emerging from surveys.

Various automated methods have been proposed to identify mergers from image surveys, all making significant contributions to the knowledge of mergers but also having some limitations.
The LDA method implemented by \cite{Nevin2019,Nevin2023} allowed us to identify a modest but statistically significant higher incidence of major mergers in our AGN samples when compared to properly matched control samples. This is a powerful method that represents an important improvement towards the identification of mergers and merger stages under a wide variety of conditions for galaxies in the local Universe, taking full advantage of the limitations of single imaging. It would be desirable to calibrate the LDA method with imaging coming from deeper surveys.

%%%%%%%%%%%%%%%%%%%%%%%%%%%%%%%%%%%%%%%%%%%%%%%%%%%

\section*{Acknowledgements}

ECS acknowledges the fellowship 825458 from CONACyT. JAVM acknowledges financial support from CONACyT grant 252531. HMHT acknowledges support from PAPIIT/UNAM grant IG101222. CAN thanks support from DGAPA-UNAM grant IN111422 and CONACyT project Paradigmas y Controversias de la Ciencia 2022-320020.

%%%%%%%%%%%%%%%%%%%%%%%%%%%%%%%%%%%%%%%%%%%%%%%%%%
\section*{Data Availability}

The data underlying this article are available at the MaNGA-Pipe3D Valued Added Catalog at \url{https://www.sdss.org/dr17/manga/manga-data/manga-pipe3d-value-added-catalog/} and \url{http://ifs.astroscu.unam.mx/MaNGA/Pipe3D_v3_1_1/tables/}. The datasets were derived from sources in the public domain using the SDSS-IV MaNGA public Data Release 15 at \url{https://www.sdss.org/dr15/}, the NASA-Sloan Atlas (NSA) catalog at \url{http://www.nsatlas.org/}, and the DESI Legacy Survey at \url{https://www.legacysurvey.org/}.



\bibliographystyle{mnras}
\bibliography{example} 
%%%%%%%%%%%%%%%%%%%%%%%%%%%%%%%%%%%%%%%%%%%%%%%%%%

%%%%%%%%%%%%%%%%%%%%%%%%%%%%%%%%%%%%%%%%%%%%%%%%%%
\appendix
\section{The flux contamination from a nuclear source and its impact on the physical properties of galaxies hosting AGNs}
\label{appendix:A}

In this Appendix we summarize our procedures to estimate the contaminant flux coming from a central AGN source after a 2D bulge/disk/bar/point-source decomposition of the $r$-band SDSS images of type I and type II AGN hosts using Galfit \citep{Peng2002,Peng2010}. To that purpose, point source function (PSF) images, masking images and pixel noise maps were generated for each galaxy. In a first step, we used as priors the results of our 1D fits from the STSDAS isophotal analysis of the Image Reduction and Analysis Facility (IRAF). Then, the S\'ersic index (n), axial ratio, and position angle were used to model a 2D generalized S\'ersic bulge and an exponential disk. In a second step, a higher order decomposition was carried out by simultaneously fitting in addition of the bulge and disk, either a Ferrers \citep{Binney1987} or a S\'ersic \citep{Geda2022} model when a bar is present. 

\begin{figure}
    \centering
    \includegraphics[width=0.9\columnwidth]{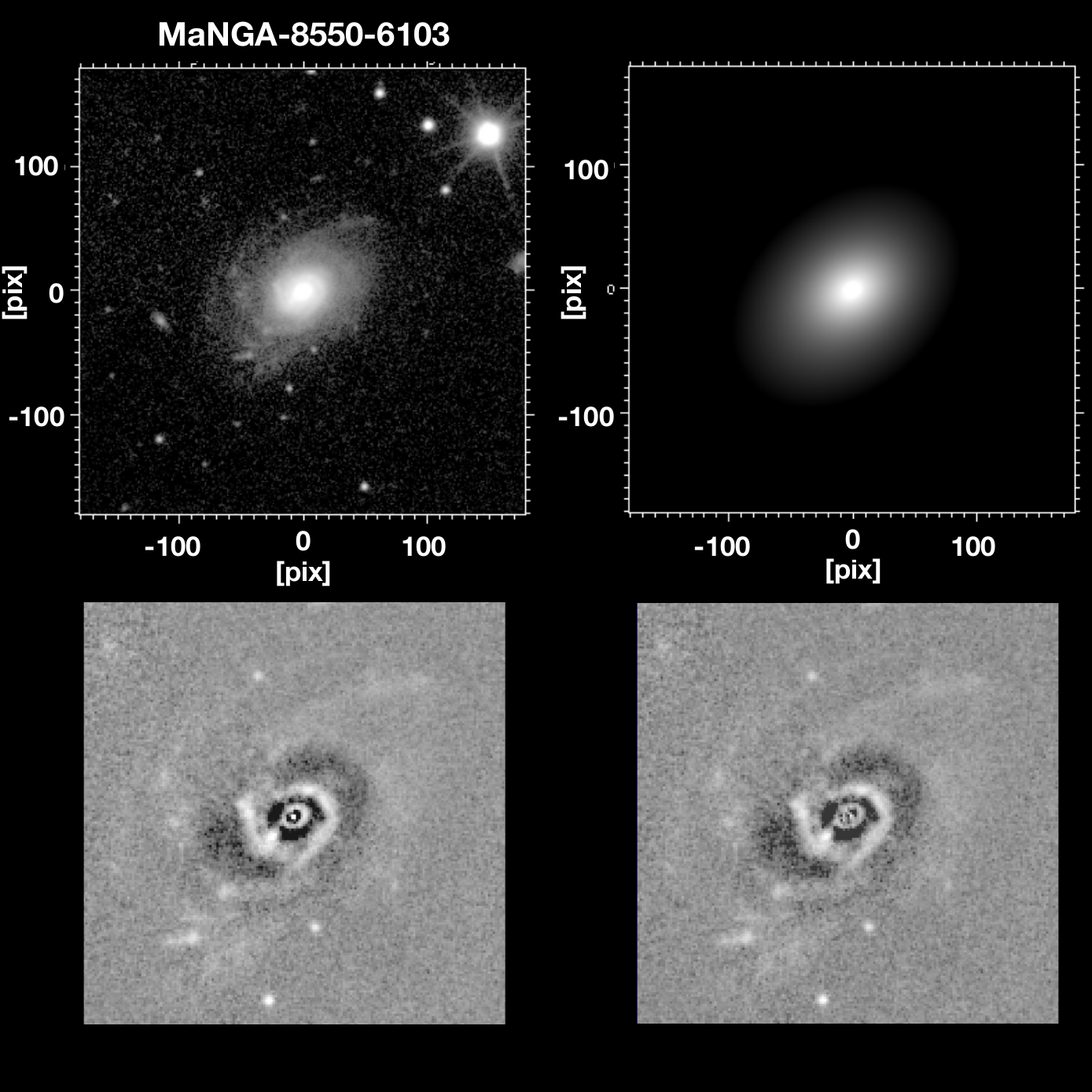}
    \caption{2D decomposition of MANGA 1-211103. Top left panel is the $r$-band SDSS image of the galaxy, top right is the best 2D model of the galaxy, bottom left and right are zooms of the residual images before and after the nuclear source subtraction.}
    \label{fig:galfit}
\end{figure}

At this stage, a careful inspection of the residual images after subtracting our best 2D model was carried out to verify for the presence of a residual flux component in the nuclear region. Then in a final step, a simultaneous fitting including a PSF, or Gaussian model is carried out. The integral flux from the model Gaussian or psf functions were interpreted as the contaminant flux coming from the central AGN source 
that can be used to correct colors, absolute magnitudes and color dependent stellar masses. Figure \ref{fig:galfit} illustrates our procedures in the case of the galaxy MaNGA 1-211103 (8550-6103).   

The residual images after our 2D decomposition, in combination with the residual images from the DESI survey catalogs and our filter-enhanced images, are very useful to provide additional evidence of inner structures like bars and outer structures like faint arms, outer rings or pseudo-rings and tidal features (see also Figure \ref{fig:mos-typeI}). 

\begin{figure}
\centering
  \includegraphics[width=\columnwidth]{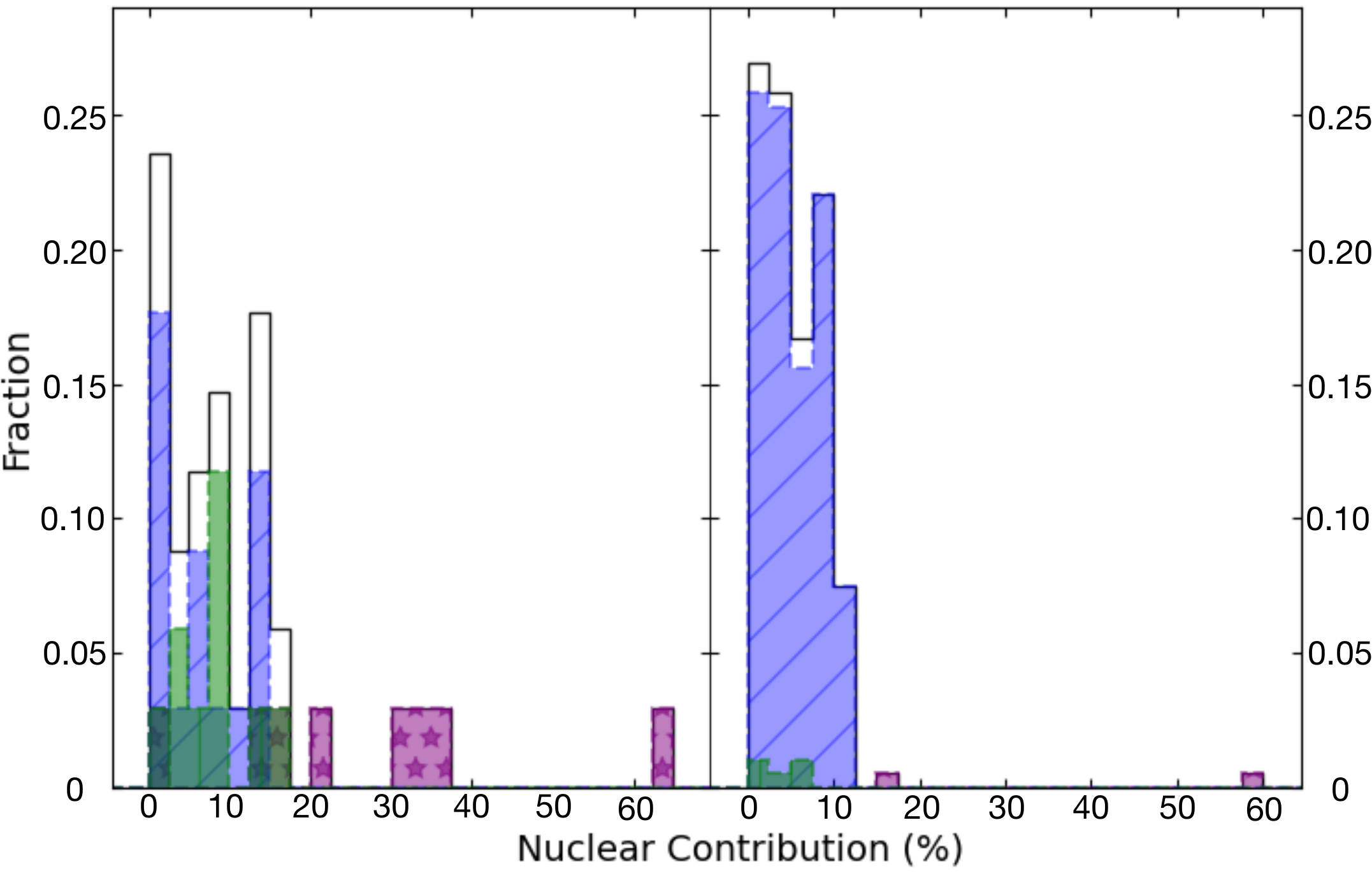}
  \vspace{-0.3cm}
  \caption{The distribution of the estimated point-source (p-s) flux contribution fraction ($f_{\rm p-s}$/$f_{\rm tot}$) coming from the nuclear AGN source in our type I AGN (left panel) and type II AGN (right panel) samples after a 2D bulge/disk/bar/point-source decomposition to the SDSS images in the $r$-band. Bar histograms are color-codded according to a classification of the central optical spectra in our sample galaxies \citep{Cortes2022}; purple for galaxies having AGN-dominated spectra, green  for galaxies having intermediate spectra and blue for galaxies having host-dominated spectra.}
\label{fig:nuc_contribt1}
\end{figure}

Figure \ref{fig:nuc_contribt1} shows the distribution of the flux contribution fraction coming from a central point source (p-s) in terms of the total flux of a host galaxy, $f_{p-s}/f_{tot}$, detected in 34 out of 47 type I AGN (left panel) and 150 out 236 type II AGN hosts (right panel). The histograms are color-coded according to a classification of the central (2.5-3 arcsec aperture) optical spectra as AGN-dominated spectra (purple), intermediate type spectra (green) and host-dominated spectra (blue) presented in \cite{Cortes2022} for type I AGN and estimated in this work for type II AGN hosts.

The histograms of Figure \ref{fig:nuc_contribt1} show that p-s nuclear contaminant fluxes higher than 20\% come only from type I and II AGN host showing central AGN-dominated spectra. In contrast, these fluxes do not exceed 20\% in the case of AGN hosts showing central spectra of the intermediate and host-dominated types.

\begin{figure}
\centering
  \includegraphics[width=\columnwidth]{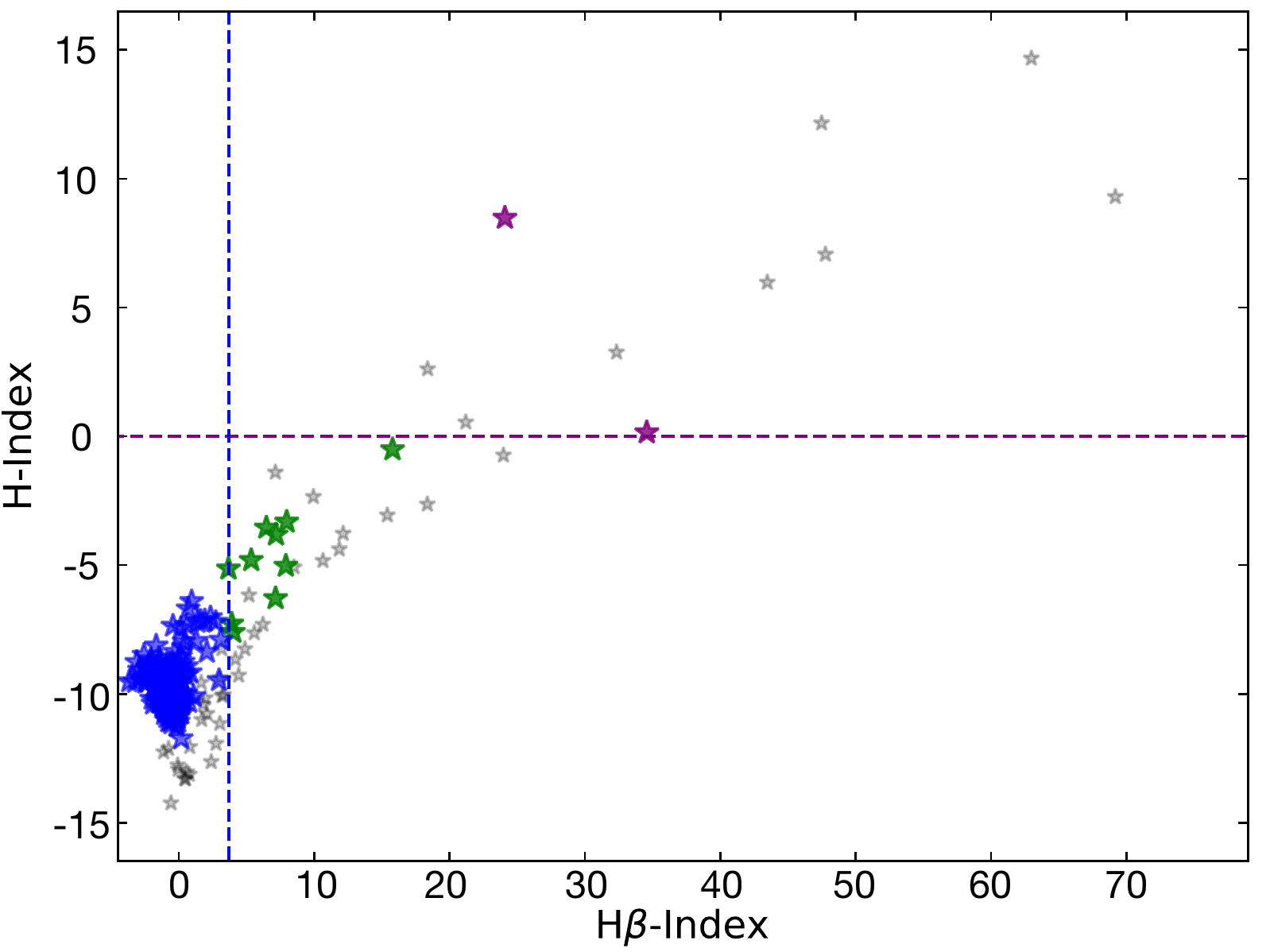}
  \vspace{-0.3cm}
  \caption{H$\beta$--H-Index diagram. The dashed lines indicate the limit regions occupied by the AGN hosts according to the measured indices in their central integrated spectra \citep[see][]{Cortes2022}. Blue stars are for host-dominated, green stars for intermediate, and purple for AGN-dominated spectra in type II AGN. Similarly, grey stars are for the type I AGN sample. }
\label{fig:indext2}
\end{figure}

\cite{Cortes2022} classified the central integrated spectra of type I AGN hosts in the MaNGA survey according to a CaII H and \hb\ indexes diagram to quantify their appearance in terms of the dominant features (host or AGN features) in the optical spectra, yielding a classification in terms of AGN-dominated, Intermediate and host-dominated spectra. The dashed blue and red lines in Figure \ref{fig:indext2} are the limit regions that best separate the H-H$\beta$ index classification of the optical spectra. The corresponding results for type II AGN hosts are also presented.

Figure \ref{fig:nuc_contribt1} and Figure \ref{fig:indext2}
suggest that a potential correlation could exist between a measure of the appearance of the central optical spectra (H-indices) and the amount of contaminant central flux ($f_{\rm p-s}$/$f_{\rm tot}$) measured in the r -band images after a 2D image decomposition. The AGN hosts with host-dominated or intermediate-type central spectra always have a central contaminant flux ($< 20\%$). Only the few cases of AGN hosts with central AGN-dominated spectra show a higher ($> 20\%$) central contaminant flux.

A classification of the optical central spectra by using the  H-indices diagram thus could be used as an empirical predictor of the amount of contaminant flux coming from a central AGN source measured in optical r -band images. Though higher numbers are desirable for a more robust analysis also in terms of different optical wavelengths, this potential correlation could be useful as a first order correction for photometric quantities like colors, absolute magnitudes and color-dependent stellar masses of large AGN samples emerging from surveys.

\subsection{Total Corrected Stellar Mass Estimates}

To estimate the total stellar masses for our type I and II sample hosts, we retrieved extinction-corrected total (S\'ersic) magnitudes and ($g-i$) colors for each galaxy coming from the NSA catalog \citep{Blanton2011}. The corresponding magnitudes are then corrected  for the presence of the nuclear contaminant flux by using the estimated $r$-band p-s flux contributions in type I and type II AGN hosts. These flux contributions were assumed as similar in the $r$ and $i$ bands to further proceed with an estimate of a decontaminated $i$ -band absolute magnitude $M_{i}$ for each galaxy.
The ($g-i$) colors come from S\'ersic magnitudes in the NSA, which were K-corrected to $z=0.0$, corrected by galactic extinction following \cite{Schlegel1998}, and by the evolution factor E($z$) following \cite{Dragomir2018}. An additional internal extinction correction was considered coming from \cite{Salim2018}.

The total stellar masses, (M$_*$), are then estimated by adopting the expression in \cite{Taylor2011}:
 
\begin{equation}
    \rm log(M_*/M\odot)=1.15+0.70(g-i)+0.40M_i
\end{equation} 

where M$_i$ is the flux-corrected corrected $i$-band absolute magnitude of a galaxy. Absolute magnitude for our galaxies were estimated by adopting h=0.7. These corrected masses thus estimated are then used all along the present paper. However, notice that these corrections do not change the main results and interpretations emerging from our analysis.

Recently, \cite{Getachew2022} quantified how the contribution of a bright nuclear (AGN) point source of different intensity could affect the values of the most commonly used non-parametric image predictors at z $\sim$ 0, finding that (i) light concentration parameters (e.g., Abraham Concentration, Gini, $M_{20}$, and Conselice Concentration) are less affected by AGN in early-type galaxies than in late-type galaxies and that (ii) the  Gini/Asym - Abraham Concentration and Conselice Concentration - $M_{20}$ diagrams are the most stable for classifying both early and late-type AGN hosts for intensities not higher than 10\% - 25\% of AGN contribution to the total optical light.  

Figure \ref{fig:nuc_contribt1} shows that the vast majority of the detected point sources in our AGN samples have intensity contribution levels lower than 20\%  implying a non-important effect of the AGN presence in our estimates of the image predictors presented in \S\S\ \ref{sec:decomposition}.

\section{Image Predictors}
\label{appendix:B}

\begin{figure}
\centering
  \includegraphics[width=\columnwidth]{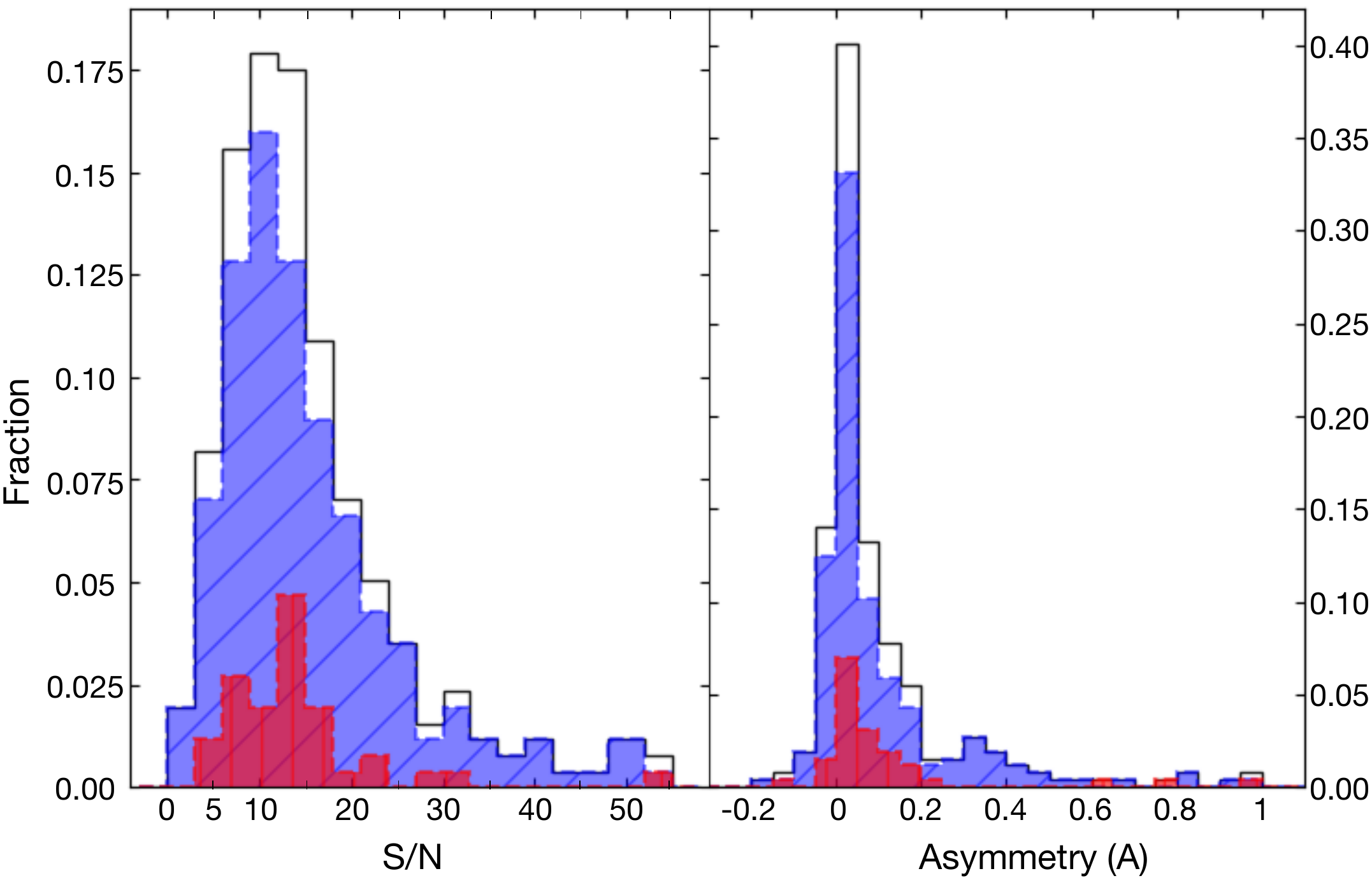}
  \vspace{-0.2cm}
  \caption{S/N distribution (left panel) and Asymmetry distribution (right panel) for type 1 AGN (red) and type 2 AGN (blue) samples.} 
  \label{fig:SN_Asy}
\end{figure}

In this Appendix we present our image predictor estimates in Table~\ref{tab:type1imaginpredictors} and Table~\ref{tab:type2imaginpredictors} and review our estimates of the asymmetry parameter looking for anomalous cases that could affect the LDA predictions.

Part of the ability of the method proposed by \cite{Nevin2019} to identify mergers relies on a correct estimate of the image predictors, particularly of the asymmetry parameter. \citet{Conselice2000} have shown that the choice of an appropriate center of rotation is an important factor to take into account. In this regard in the present work the center of rotation is measured through an iterative process that finds the minimum asymmetry following \cite{Conselice2000} and \cite{Lotz2008}.

Other important factors that could alter the measured asymmetry and also of clumpiness predictors are the presence of a significant nonzero background level and lower S/N ratios. 

\cite{Nevin2019} experimented with their simulated images to delimit the range surface brightness that can return a consistent merger classifications, finding that it is closely tied to the behavior of the imaging predictors, which are sensitive to resolution and  average S/N per pixel (<S/N>). \cite{Lotz2004} found that the Gini, M20, C, A, and S image predictors are reliable to 10\% for <S/N> $>$ 2 decreasing systematically with <S/N> below this level. \cite{Nevin2019} implemented a <S/N> cutoff of 2.5 (for all pixels within a segmentation map) to avoid that the imaging predictors, particularly A and S, become unreliable or could decrease to negative values below this threshold. 

The left panel of Figure \ref{fig:SN_Asy} shows the distribution of mean <S/N> ratios above which the image predictors were estimated in our type I and II AGN samples. In addition, the right panel shows the distribution of asymmetry values. The incidence of <S/N> $\leq$ 2.5 values is less than $\sim$2\% for the combined type I and II AGN samples thus making the choice of <S/N> $\geq$ 2.5 threshold an appropriate cutoff for a correct estimate of the image predictors in our samples. 

In practice the asymmetry predictor, A, uses a correction factor \citep[e.g.,][]{Conselice2000} from a background image. Notice however that the background sky levels can vary across and between the frames, leading in some cases to overcorrections in the asymmetry values and thus yielding negative values \citep[e.g.,][]{Lotz2004,Giese2016,Thorp2021}.

The right panel of Figure \ref{fig:SN_Asy} shows that the incidence of negative asymmetry values A $<$ 0 in our combined type I and II AGN samples is 14\%\ (10\% having only slightly negative values  $-0.05 < A < 0$ and 4\% having a strong over-correction $A > -0.05$). We have verified the impact of those negative 
values on our results, finding 5 galaxies with negative asymmetry values in the type I AGN sample, but only one, 1-52660, classified as major merger (see Table \ref{tab:type1imaginpredictors}). In the type II AGN sample, there are 40 galaxies with negative asymmetry values, but only two 1-277257 and 1-314700 classified as major mergers (see Table \ref{tab:type2imaginpredictors}). These galaxies have been omitted from the merger fraction analysis (Section 4.1). However, their omission do not produce significant changes in our final results. 
 
The reader may find useful to take into account the flagging codes in the last two columns of Table~\ref{tab:type1imaginpredictors} and Table~\ref{tab:type2imaginpredictors} indicating the 
<S/N> ratios above which the image predictors were estimated as well as the negative asymmetry values (flag = 1).

\begin{table*}
\begin{center}
\begin{tabular}{cccccccccc}
\hline
\hline
\textbf{MANGAID}	&	\textbf{Gini}	&	\textbf{M$_{20}$}	&	\textbf{Concentration (C)}	&	\textbf{Asymmetry (A)}	&	\textbf{Clumpiness (S)}	&	\textbf{Sersic (n)}	&	\textbf{Shape Asymmetry (A$_S$)}	&	\textbf{S/N} & \textbf{Negative A}	\\
(1)	&	(2)	&	(3)	&	(4)	&	(5)	&	(6)	&	(7)	& 	(8)	& 	(9) & (10)	\\
\hline	\\																
1-113712	&	0.7364	&	-0.6968	&	3.3776	&	0.7714	&	0.2791	&	0.8846	&	0.3762	& 	6.5	& 0\\
1-180204	&	0.4621	&	-1.2940	&	2.3800	&	0.1105	&	0.0368	&	0.4337	&	0.1315	& 	8.7	& 0\\
1-113405	&	0.4959	&	-1.0232	&	1.9382	&	0.9645	&	0.0000	&	0.9606	&	0.9071	& 	183.4	& 0\\
1-596598	&	-	&	-	&	-	&	-	&	-	&	-	&	-	& 	-	& -\\
1-24092	&	0.5496	&	-1.7590	&	2.8795	&	0.0510	&	0.0000	&	1.4835	&	0.6206	& 	87.2	& 0\\
1-24148	&	0.5239	&	-2.0122	&	3.4040	&	0.0170	&	0.0298	&	0.7034	&	0.1004	& 	32.7	& 0\\
1-548024	&	0.6229	&	-2.3949	&	4.2029	&	0.0920	&	0.0172	&	1.9383	&	0.3602	& 	5.1	& 0\\
1-43214	&	0.5319	&	-1.6577	&	2.6976	&	0.0660	&	0.0000	&	1.4200	&	0.5901	& 	114.7	& 0\\
1-121075	&	0.5975	&	-2.0788	&	3.8073	&	0.0056	&	0.0071	&	1.6886	&	0.2522	& 	15.4	& 0\\
1-52660	&	0.5937	&	-2.3476	&	4.2695	&	-0.1284	&	-0.1833	&	1.8635	&	0.6064	& 	3.1	& 1\\
1-460812	&	0.9352	&	-2.0906	&	3.6974	&	0.0815	&	0.0123	&	1.2318	&	0.1256	& 	23.4	& 0\\
1-523004	&	0.4964	&	-1.0691	&	3.1412	&	0.1069	&	0.0422	&	0.6471	&	0.1689	& 	15.0	& 0\\
1-235576	&	0.5927	&	-2.2961	&	4.1975	&	0.0572	&	0.0060	&	1.1765	&	0.1101	& 	13.1	& 0\\
1-620993	&	0.5014	&	-1.7753	&	2.9202	&	0.1525	&	0.0332	&	0.6023	&	0.0601	& 	14.6	& 0\\
1-418023	&	0.5234	&	-1.8563	&	3.0080	&	0.0151	&	0.0565	&	1.2239	&	0.0549	& 	52.7	& 0\\
1-256832	&	0.6238	&	-1.6832	&	3.8581	&	0.1790	&	0.0323	&	1.9085	&	0.1733	& 	12.2	& 0\\
1-593159	&	0.6103	&	-2.4581	&	4.1821	&	0.0464	&	0.0176	&	2.1388	&	0.1036	& 	11.4	& 0\\
1-210017	&	0.5324	&	-2.1717	&	3.9716	&	0.1179	&	0.0218	&	1.2150	&	0.4347	& 	12.2	& 0\\
1-90242	&	0.4971	&	-1.6246	&	2.7102	&	0.0511	&	0.0000	&	2.6333	&	0.0947	& 	303.3	& 0\\
1-90231	&	-	&	-	&	-	&	-	&	-	&	-	&	-	& 	-	& -\\
1-594493	&	0.5349	&	-1.4296	&	3.0495	&	0.2143	&	0.0016	&	2.5344	&	0.0490	& 	13.1	& 0\\
1-95585	&	0.5230	&	-2.1264	&	3.2352	&	0.0439	&	0.0069	&	0.8587	&	0.4448	& 	7.7	& 0\\
1-550901	&	0.6339	&	-0.7709	&	4.0144	&	0.1937	&	0.0671	&	1.6806	&	0.3342	& 	15.2	& 0\\
1-71974	&	0.6214	&	-2.8594	&	5.4145	&	0.1014	&	0.0019	&	2.0906	&	0.1908	& 	7.4	& 0\\
1-604860	&	0.6037	&	-2.3170	&	4.1363	&	-0.0173	&	0.0014	&	1.9112	&	0.1124	& 	14.6	& 1\\
1-44303	&	-	&	-	&	-	&	-	&	-	&	-	&	-	& 	-	& -\\
1-574519	&	0.5750	&	-2.1095	&	3.5669	&	0.0148	&	0.0136	&	1.3262	&	0.0942	& 	9.2	& 0\\
1-163966	&	0.6208	&	-2.4118	&	4.1332	&	0.0339	&	0.0153	&	1.3991	&	0.0589	& 	16.7	& 0\\
1-94604	&	0.5488	&	-2.1584	&	3.6349	&	0.0093	&	-0.0020	&	1.3831	&	0.0869	& 	13.7	& 0\\
1-423024	&	0.5178	&	-1.9870	&	3.0941	&	0.0428	&	0.0069	&	0.9810	&	0.0736	& 	13.9	& 0\\
1-174631	&	0.5503	&	-2.1929	&	3.8326	&	0.0477	&	0.0110	&	1.6338	&	0.1069	& 	10.4	& 0\\
1-149561	&	0.5562	&	-2.0755	&	3.4164	&	-0.0168	&	0.0175	&	1.2032	&	0.0885	& 	14.4	& 1\\
1-614567	&	0.6446	&	-1.1382	&	2.1943	&	0.6284	&	0.1157	&	1.6698	&	0.6389	& 	10.2	& 0\\
1-210186	&	0.4477	&	-1.9328	&	2.8121	&	0.0342	&	0.0139	&	0.4350	&	0.0376	& 	10.7	& 0\\
1-295542	&	0.6063	&	-2.0405	&	3.7985	&	0.0201	&	-0.0074	&	2.0132	&	0.1038	& 	27.0	& 0\\
1-71872	&	0.5704	&	-2.4916	&	4.1627	&	-0.0272	&	-0.0945	&	4.0290	&	0.0388	& 	7.7	& 1\\
1-71987	&	0.5560	&	-2.0741	&	3.5306	&	0.0168	&	0.0072	&	1.4681	&	0.4039	& 	14.3	& 0\\
1-37863	&	0.6141	&	-2.2581	&	4.4121	&	0.1124	&	0.0138	&	2.3847	&	0.3035	& 	13.7	& 0\\
1-37385	&	0.5861	&	-2.5195	&	4.3766	&	0.0032	&	-0.0530	&	2.4755	&	0.3449	& 	4.8	& 0\\
1-37336	&	0.5931	&	-2.1943	&	3.7534	&	-0.0216	&	0.0029	&	1.5568	&	0.0860	& 	8.9	& 1\\
1-37633	&	0.6196	&	-2.2676	&	3.9954	&	0.0014	&	0.0177	&	2.0040	&	0.1216	& 	19.2	& 0\\
1-24660	&	0.5560	&	-1.9830	&	3.2225	&	0.0394	&	-0.0051	&	1.0989	&	0.0657	& 	8.9	& 0\\
1-574506	&	0.5404	&	-2.0429	&	3.5252	&	0.0211	&	0.0088	&	1.3389	&	0.1329	& 	21.4	& 0\\
1-574504	&	0.5669	&	-2.4553	&	4.3214	&	0.0889	&	0.0179	&	1.0877	&	0.2836	& 	17.6	& 0\\
1-298111	&	-	&	-	&	-	&	-	&	-	&	-	&	-	& 	-	& -\\
1-385623	&	0.5744	&	-1.7129	&	3.0685	&	0.0760	&	0.0000	&	3.2054	&	0.0648	& 	121.9	& 0\\
1-523211	&	0.5943	&	-2.2527	&	3.9481	&	0.0002	&	0.0122	&	1.0985	&	0.1303	& 	13.9	& 0\\
\hline	
\hline
\end{tabular}
\end{center}
\caption{Main values of the imaging predictors for the 47 galaxies with type I active nuclei according to our selection criteria \citep{Cortes2022}. These seven values were used to estimate the LD1 parameter to classify the merging galaxy following \citet{Nevin2019}.}
\label{tab:type1imaginpredictors}
\end{table*}

\begin{table*}
\begin{center}
\begin{tabular}{cccccccccc}
\hline
\hline
\textbf{MANGAID}	&	\textbf{Gini}	&	\textbf{M$_{20}$}	&	\textbf{Concentration (C)}	&	\textbf{Asymmetry (A)}	&	\textbf{Clumpiness (S)}	&	\textbf{Sersic (n)}	&	\textbf{Shape Asymmetry (A$_S$)}	&	\textbf{S/N} & \textbf{Negative A}	\\
(1)	&	(2)	&	(3)	&	(4)	&	(5)	&	(6)	&	(7)	& 	(8)	& 	(9) & (10)	\\
\hline	\\																
12-84677	&	0.5604	&	-2.1705	&	3.5267	&	0.0272	&	0.0089	&	1.2985	&	0.1369	& 	12.9	& 0\\
1-113651	&	0.5320	&	-2.0499	&	3.3046	&	-0.0235	&	0.0129	&	0.9991	&	0.0274	& 	10.4	& 1\\
1-547210	&	0.6707	&	-0.4428	&	4.5952	&	-0.1415	&	0.0629	&	0.9632	&	0.4015	& 	5.1	& 1\\
1-547402	&	0.5764	&	-1.8962	&	3.7603	&	-1.4056	&	-0.3170	&	0.9121	&	0.2598	& 	1.5	& 1\\
1-177493	&	0.6861	&	-0.4645	&	3.2641	&	0.1041	&	0.0183	&	0.9164	&	0.2112	& 	11.1	& 0\\
1-177528	&	0.4673	&	-2.0513	&	3.1695	&	-0.0392	&	0.0010	&	0.5357	&	0.1092	& 	12.8	& 1\\
1-180629	&	0.4731	&	-1.7810	&	2.8584	&	0.0022	&	0.0260	&	0.9227	&	0.1062	& 	6.6	& 0\\
1-180537	&	-	&	-	&	-	&	-	&	-	&	-	&	-	& 	-	& -\\
1-25554	&	0.5502	&	-2.2134	&	3.6783	&	-0.0956	&	-0.0334	&	1.3119	&	0.1776	& 	7.8	& 1\\
...    	&	...	&	...	&	...	&	...	&	...	&	...	&	...	& 	& ...	\\
\hline	
\hline
\end{tabular}
\end{center}
\caption{Same as \ref{tab:type1imaginpredictors} but for the 236 galaxies with type II active nuclei. The complete data will be available in the online version.}
\label{tab:type2imaginpredictors}
\end{table*}

% Don't change these lines
\bsp	% typesetting comment
\label{lastpage}
\end{document}